\documentclass[pdflatex,sn-mathphys-num]{sn-jnl}

\usepackage{graphicx}%
\usepackage{multirow}%
\usepackage{amsmath,amssymb,amsfonts}%
\usepackage{amsthm}%
\usepackage{mathrsfs}%
\usepackage[title]{appendix}%
\usepackage{xcolor}%
\usepackage{textcomp}%
\usepackage{manyfoot}%
\usepackage{booktabs}%
\usepackage{algorithm}%
\usepackage{algorithmicx}%
\usepackage{algpseudocode}%
\usepackage{listings}%

\usepackage{tabularx}
\usepackage{url}
\usepackage{xurl}
\usepackage{ragged2e}
\usepackage{enumitem}

\theoremstyle{thmstyleone}%
\theoremstyle{thmstyletwo}%

\theoremstyle{thmstylethree}%

\begin{document}

\title[Aurora]{Aurora: Architecting Argonne’s First Exascale Supercomputer for Accelerated Scientific Discovery}

\author[1]{\fnm{William E.} \sur{Allcock}}
\author[1]{\fnm{Benjamin S.} \sur{Allen}}
\equalcont{Authors are listed alphabetically.}
\author[2]{\fnm{James} \sur{Anchell}}
\author[1]{\fnm{Victor} \sur{Anisimov}}
\author[1]{\fnm{Thomas} \sur{Applencourt}}
\author[1]{\fnm{Abhishek} \sur{Bagusetty}}
\author[1]{\fnm{Ramesh} \sur{Balakrishnan}}
\author[1]{\fnm{Riccardo} \sur{Balin}}
\author[1]{\fnm{Solomon} \sur{Bekele}}
\author[1]{\fnm{Colleen} \sur{Bertoni}}
\author[1]{\fnm{Cyrus} \sur{Blackworth}}
\author[2]{\fnm{Renzo} \sur{Bustamante}}
% ...

\affil*[1]{\orgname{Argonne National Laboratory}, \orgaddress{\city{Lemont}, \state{Illinois}, \country{USA}}}
\affil*[2]{\orgname{Intel Corporation}, \orgaddress{\city{Santa Clara}, \state{California}, \country{USA}}}
\affil*[3]{\orgname{Hewlett Packard Enterprise}, \orgaddress{\city{Spring}, \state{Texas}, \country{USA}}}

\abstract{Aurora is Argonne National Laboratory's pioneering Exascale supercomputer, designed to accelerate scientific discovery with cutting-edge architectural innovations. Key new technologies include the Intel® Xeon® Max Series (code-named Sapphire Rapids) with support for High Bandwidth Memory (HBM), alongside the Intel® Data Center GPU Max Series (code-named Ponte Vecchio) on each compute node. Aurora also integrates the Distributed Asynchronous Object Storage (DAOS), a novel exascale storage solution, and leverages Intel's oneAPI programming environment. This paper presents an in-depth exploration of Aurora’s node architecture, the HPE Slingshot interconnect, the supporting software ecosystem, and DAOS. We provide insights into standard benchmark performance and applications readiness efforts via Aurora’s Early Science Program and the Exascale Computing Project.}



\begin{titlepage}
    \maketitle


\section*{Acknowledgments}\label{Acknowledgment}

This research was supported by and used resources of the Argonne Leadership Computing Facility, which is a DOE Office of Science User Facility supported under Contract DE-AC02-06CH11357.
This work is associated with an ALCF Aurora Early Science Program project.
We would like to thank all the ESP postdocs who helped advance their projects and the Program en route to their next career stages. Some have continued at Argonne and are co-authors on this paper; thank you to the others: Denis Boyda, Raghuveer Chimata, Raymundo Hernandez Esparza, Jun Fang, William Huhn (term staff), Brian MacKie-Mason, Geng Liu, Pankaj Rajak, Jonathan Thirman, Rafael Vescovi, and Azton Wells.
All Argonne authors were supported by the Office of Science, U.S. Department of Energy, under contract DE-AC02-06CH11357.
Some of this work was done on a pre-production supercomputer with early versions of the Aurora software development kit.

The submitted manuscript has been created by UChicago Argonne, LLC, Operator of Argonne National Laboratory (“Argonne”). Argonne, a U.S. Department of Energy Office of Science laboratory, is operated under Contract No. DE-AC02-06CH11357. The U.S. Government retains for itself, and others acting on its behalf, a paid-up nonexclusive, irrevocable worldwide license in said article to reproduce, prepare derivative works, distribute copies to the public, and perform publicly and display publicly, by or on behalf of the Government. The Department of Energy will provide public access to these results of federally sponsored research in accordance with the DOE Public Access Plan. http://energy.gov/downloads/doe-public-access-plan.

\end{titlepage}


\section{Introduction}\label{Introduction}

\begin{figure}[htbp]
    \centering
    \includegraphics[width=\linewidth]{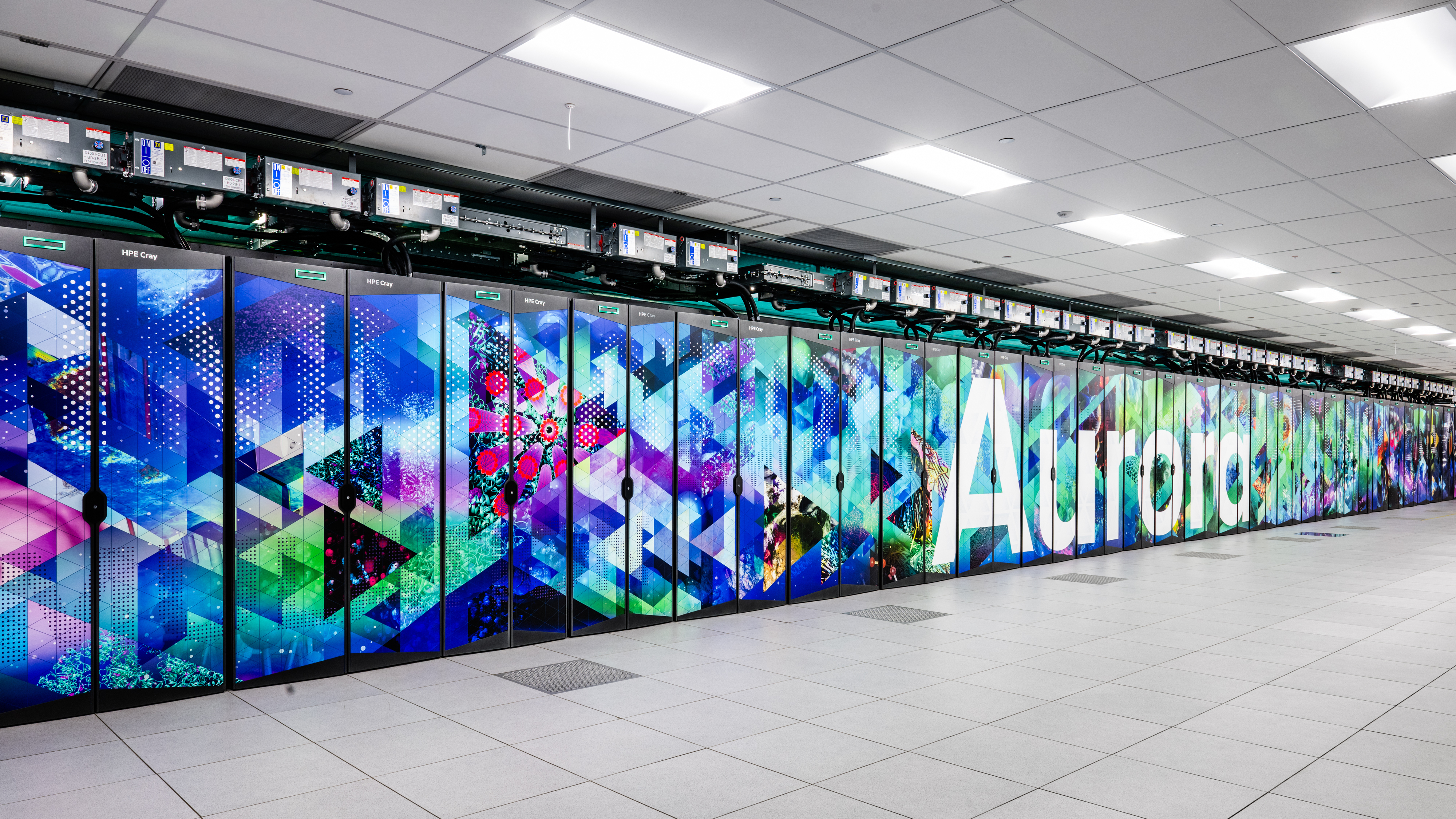}
    \caption{Aurora's first row of cabinets at the ALCF.}
    \label{fig:aurora-image}
\end{figure}

Aurora is an exascale-class supercomputer hosted at the Argonne Leadership Computing Facility (ALCF), shown in Figure~\ref{fig:aurora-image}. It is among the most powerful systems built~\cite{TOP500}, representing a major step forward in computational capabilities. 
ALCF's mission to support all three pillars of science applications, i.e. Simulation, Data Science, and Machine Learning, influenced many of the design goals for Aurora. Building an exascale system of this complexity required addressing a variety of technical challenges and making informed decisions across both hardware and software components. This paper outlines the key design choices behind Aurora’s architecture and software stack, and discusses how these decisions affect system performance and scientific application workflows.


Aurora uses a new GPU architecture from Intel Corporation, the Intel Data Center GPU Max 1550, code-named \textit{Ponte Vecchio (PVC)}, which consists of over 100B transistors across many silicon tiles and process nodes~\cite{9731673}. The Xe-cores or execution units contain dual fp64 pipelines and a systolic array for low precision operations that aims to tackle both HPC and AI/ML workloads. The Intel Xeon Max series multi chiplet design~\cite{9731107} with its large core count, Advanced Matrix eXtension (AMX), and onboard high-bandwidth memory (HBM) aids in designing a system that can break through exascale for science applications. 
Aurora has the largest~\cite{largest_ss} deployment of Slingshot-11 interconnect in a dragonfly topology with over 300K fabric ports. Table~\ref{tab:aurora_compute_spec} presents the system's aggregate specifications. In addition a novel 260~PB Distributed Asynchronous Object Store (DAOS)~\cite{hennecke2020daos} that uses high performance flash with a peak bandwidth of 31~TB/s acts as the storage subsystem.

%



\begin{table}[htbp]
\centering
\caption{Aurora Aggregate Specifications.}
\begin{tabular}{lc}
\toprule
Nodes & 10,624 \\
No. of CPUs & 21,248 \\
No. of GPUs & 63,744 \\
DDR5 Memory Capacity & 10.62~PB \\    
DDR5 Memory Bandwidth & 5.31~PB/s \\  
HBM2e Memory Capacity & 9.52~PB \\    
HBM2e Memory Bandwidth & 147.46~PB/s \\  
Injection Bandwidth & 2.12~PB/s \\    
Global Bandwidth & 1.37~PB/s \\ 
\bottomrule
\end{tabular}
\label{tab:aurora_compute_spec}
\end{table}




Application readiness is demonstrated with 19 ALCF Early Science Projects~\cite{Kim_2018,Uintah,XGC1,XGC2,XGC3,MILC,Chroma,Grid,QUDA,HABIB201649,PHASTA1,PHASTA2,NAMD,nwchemex,BYLASKA2024518,FastCaloSim,MadGraph,cfdml,HARVEY,LQCDML,connectomics,BerkeleyGW,SST} and 15 ECP Applications Development (AD) projects.
In the following, Section \ref{SystemDesign} delves into the hardware design of Aurora, and Section \ref{SoftwareAndProgrammingEnvironnment} discusses the software ecosystem developed for it. 
Section \ref{EarlyApplications} explores the variety of scientific applications that are currently running on Aurora. 
This is followed by an assessment of the theoretical and actual performance characteristics of the machine in Section~\ref{PerformanceEvaluation}. 
The paper ends with a summary of key challenges (Section~\ref{Challenges}) and conclusions (Section~\ref{Conclusion}).

\section{System Design} \label{SystemDesign}

This section describes Aurora's compute blade architecture, the Slingshot 11 interconnect, and the I/O subsystem.

\subsection{Exascale Compute Blade (ECB)} \label{ECB}

The Aurora compute node, referred to as the Exascale Compute Blade (ECB), consists of two (2) Intel Xeon Max Series CPU with HBM, code-named \textit{Sapphire Rapids (SPR)}, and six (6) Intel Data Center Max GPUs, code-named \textit{Ponte Vecchio (PVC)}, for its compute components. It contains eight (8) network interface controllers (NIC) to connect the nodes together across a high-performance fabric, HPE's Slingshot. 
The 2:6 CPU-to-GPU node ratio was determined to maximize performance and density within the power envelope and physical dimensions of the ECB. 
The blade provides an aggregate measured compute performance of 145~TF/s double precision in single node HPL within a 4~kW power envelope. 


\begin{figure}[htb]
    \centering
    \includegraphics[width=0.82\linewidth]{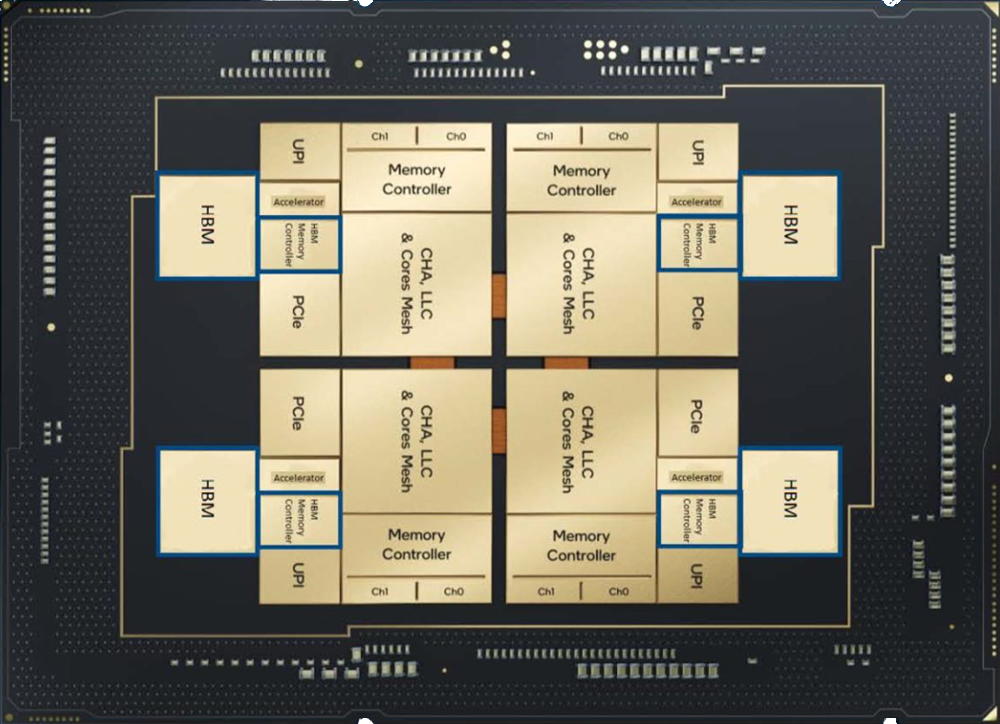}
    \caption{Intel® Xeon Max Series CPU with HBM.}
    \label{fig:spr-cpu}
\end{figure}

\subsubsection{Intel® Sapphire Rapids (SPR) CPU} The product version used on Aurora consists of 52 cores in a dual socket configuration. The underlying CPU design~\cite{9731107} is composed of four compute chiplets connected together with Intel's embedded multi-die interconnect bridge (EMIB) interface. This SPR CPU was chosen specifically for Aurora with two important changes that came from early architecture modeling of the system (as shown in Figure~\ref{fig:spr-cpu}). First, the Extreme Core Count (XCC) allowed the host CPU to be capable of moving data across the five x16 Gen5 PCIe lanes between the GPUs and NICs, in addition to being able to drive work into the various GPUs. Second, the onboard 64~GB HBM2e memory can act as a high speed buffer for staging and preprocessing data for the application. The CPU can also be booted into a different cache configuration where the HBM can act as a direct mapped cache becoming a transparent memory tier to DDR~\cite{ibeid2025hbm}. The SPR CPU also contains various accelerators, AMX in particular which can speedup low precision dense matrix operations that machine learning frameworks running on Aurora benefit from~\cite{amx}.


\begin{figure}[htbp]
    \centering
    \includegraphics[width=0.83\linewidth]{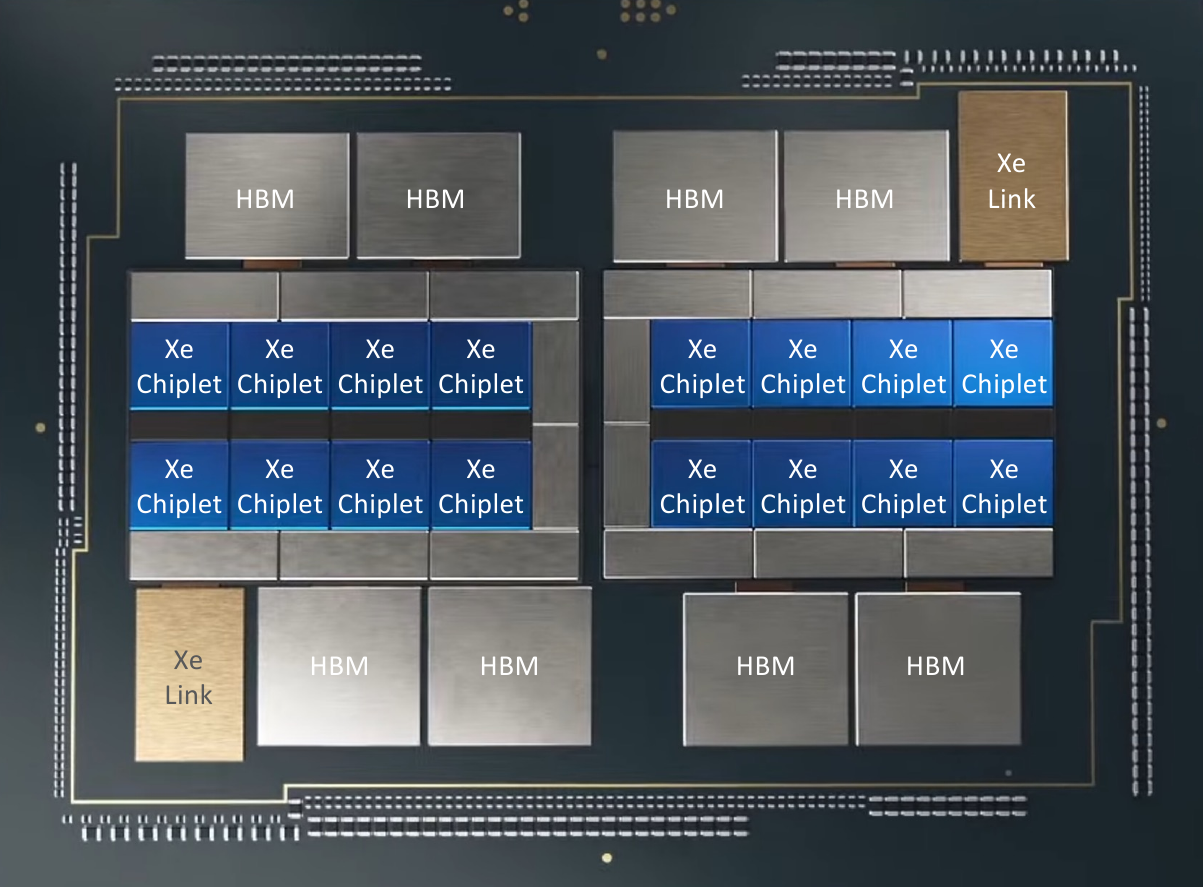}
    \caption{Intel® Data Center GPU Max.}
    \label{fig:pvc-gpu}
\end{figure}

\subsubsection{Intel Data Center GPU Max 1550 (PVC)}
PVC GPU~\cite{hotchips2021blythe, hotchips2022jiang,9731673} was designed from the outset to meet the specific exascale HPC application requirements of the Aurora project which anticipated the convergence of AI and HPC. 
PVC features a hierarchical architecture that scales effectively from previous generations of low-power integrated Intel GPUs~\cite{intelgen11gfx, intelxehpg}. The fundamental building block of PVC is the Xe-Core, which consists of eight vector and matrix engines, with a combined register file size of 512 KB. The vector unit is 512 bits wide (8-wide for double precision floating point (FP64)) and supports fused multiply-add (FMA) operations. 
The matrix unit is designed to support mixed-precision matrix multiplications, commonly used in AI applications. It is expanded to a width of 4096 bits to accommodate the ever-growing computational demands of AI. The input data formats include 32-bit tensor floating-point (TF32), 16-bit floating-point (BF16 and FP16), and integer formats (INT8 and INT4).
The register file is shared between the vector and matrix units, utilizing a common data layout to eliminate data marshaling overhead. 
It can be partitioned among hardware threads in two configurations to optimize for latency and throughput: either 8 active hardware threads with 128 512-bit registers each, or 4 active hardware threads with 256 registers each. Each vector engine can co-issue two 8-wide FP64 FMAs per clock cycle, allowing all vector engines within a Xe-Core to perform 256 FP64 operations per clock (8 vector engines/Xe-Core \(\times\) 8 (512 bits SIMD)\(\times\)2 (FMA)\(\times\)2). The decision to maintain double-rate FP64 on the vector unit, rather than the matrix unit, benefits broader range of HPC applications.

The engines within the Xe-Core share a large 512 KB high-bandwidth L1 data cache, capable of 512 bytes per clock load or 256 bytes per clock store. The L1 cache can be partially configured as highly banked Shared Local Memory, serving as a software-managed scratch pad with high scatter/gather throughput. The Xe-Core configuration in PVC is approximately twice the size of those in other contemporary GPUs, bringing memory closer to compute to minimize cache bandwidth demand. For workloads that benefit from memory-tiling optimizations, such as matrix multiplications, this design reduces bandwidth requirements by a factor of $\sqrt{2}$. The GPU clock speed can reach up to 1.6 GHz, adjustable to adhere to Thermal Design Power (TDP) constraints of the platform.
PVC groups 64 Xe-Cores into one Xe-Stack which share the Last Level Cache (LLC), HBM2e memory controllers, high-speed Stack-to-Stack interconnect within the same silicon substrate, and the Xe-link high-speed coherent fabric for remote GPU-to-GPU communication. The LLC in each Xe-Stack is a 192 MiB memory-side cache connected to its dedicated HBM2e memory stacks, providing a 64-byte interface to each Xe-Core.

Two Xe-Stacks, along with their local HBM2e memory stacks, are integrated into an Intel GPU package, as shown in Figure~\ref{fig:pvc-gpu}, totaling 128 Xe-Cores and enabling 32,768 double precision and single precision floating-point operations per clock. Only the first Xe-Stack includes the PCIe link for host connection. Data movement from the second Xe-Stack traverses the high-speed Stack-to-Stack interconnect and the PCIe link before reaching the host. The Intel PVC GPU package supports PCIe Gen5 link speeds.

\subsubsection{HPE Cassini NIC}
Cassini is a 200~Gbps HPC NIC ASIC chip developed by HPE. The host interface is PCIe Gen4 with extended speed mode (where supported by the CPU or GPU). The network link port conforms to the 200~Gbps (4×50~Gbps PAM 4) and 100~Gbps (4×50~Gbps NRZ) Ethernet standards. The link between a Cassini NIC and a Rosetta switch can carry Ethernet frames, v4 and v6 IP traffic, MPI messages, one-sided operations, low latency barrier operations, and RoCE storage traffic simultaneously.

\subsubsection{Connectivity plane}
\begin{figure}[htbp]
    \centering
    \includegraphics[width=0.85\linewidth]{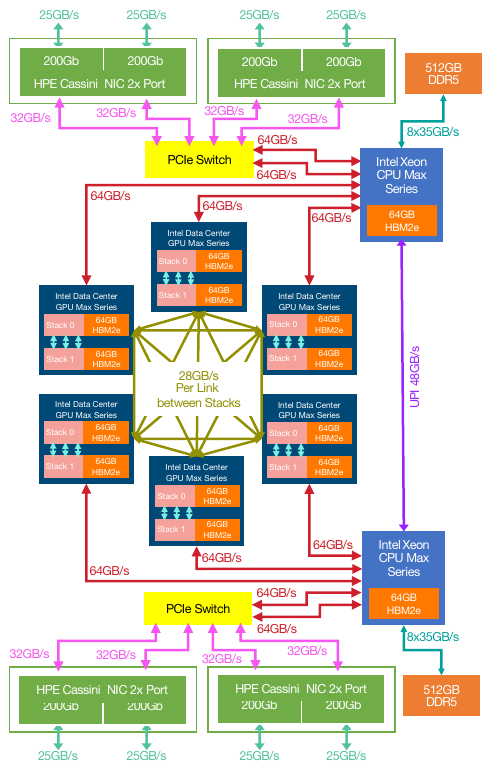}
    \caption{Aurora Exascale Compute Blade (ECB).}
    \label{fig:aurora-node}
\end{figure}

\begin{figure}[htbp]
    \centering
    \includegraphics[width=0.9\linewidth]{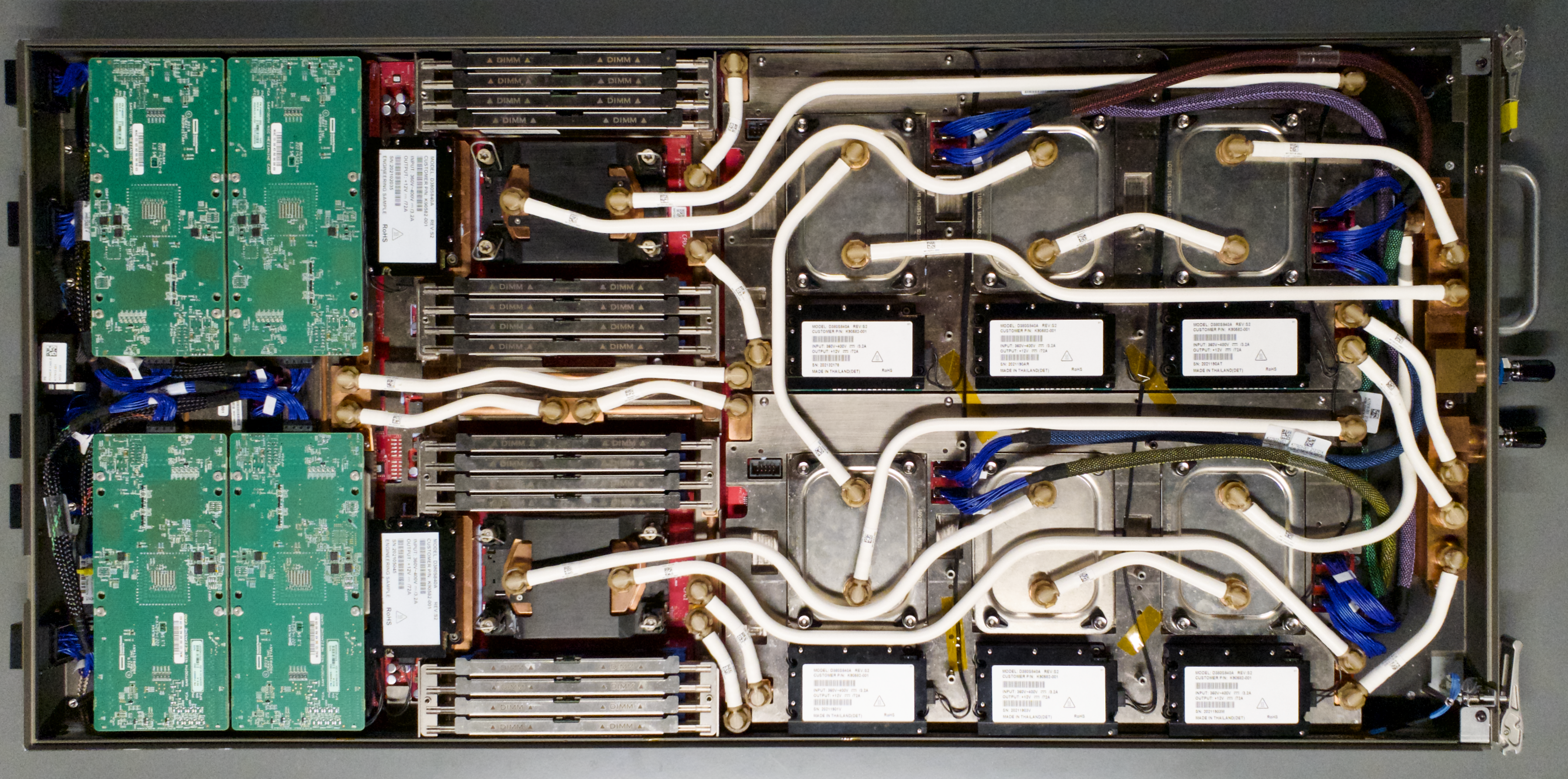}
    \caption{Physical Aurora Exascale Compute Blade (ECB).}
    \label{fig:aurora-phy-node}
\end{figure}

Figure~\ref{fig:aurora-node} illustrates the overall blade architecture and Figure~\ref{fig:aurora-phy-node} is a photograph of the blade. The Intel GPUs are directly connected to the CPUs via PCIe Gen5 x16 links offering 64~GB/s design bandwidth. The Intel® GPUs construction utilizes dual compute stack architecture and therefore results in a dual plane configuration where stacks from each GPU are connected together in all-to-all manner with dedicated Xe-Links each capable of 28~GB/s of theoretical bandwidth.


NICs are connected to the CPUs via PCIe switch that splits 2x PCIe Gen5 x16 link to 4x PCIe Gen4 x16 link. GPU memory buffers can be sent directly to the NIC through dma-buf subsystem in the Linux kernel~\cite{dma-buf}. The actual data movement happens through PCIe Peer-to-Peer DMA transfer~\cite{pcie-p2p}. This optimization allows data transfer to bypass CPU memory. MPICH can also pipeline data via host memory. This is controlled using environment variables.


\subsubsection{ECB power distribution}
Rectifiers placed in the chassis convert the AC power to 380V DC and to 12V standby. The 380V DC power is then routed to the intermediate voltage converter (IVOC) on the ECB via the Mid Plane Connector. The IVOC then converts the 380V to 12V level required for the board voltage regulator components. There are 8 IVOCs in the ECB, one for each of the 6 GPUs and one for each of the two CPUs. ECB offers sustained power throttling capabilities to the CPU and GPU via both inband and out-of-band mechanisms. Further, power bias can also be programmed to shift power between the CPUs and GPUs. ECB was designed to operate at sustained power draw of 4~kW and a peak power draw for shorter duration ($\approx$5 to 20ms) at 4.6~kW. This is designed to support the GPU's short duration power draw which can be higher than their sustained design power of 500~W. In nominal workload execution the ECB operates at ~3.8~kW of power draw. The CPUs draw approximately 350~W each and the GPUs draw 500~W each when workloads actively use them.


\subsection{Scale Out Design}

\subsubsection{Slingshot-11 Interconnect} Aurora uses the HPE Shasta Hardware Architecture~\cite{shasta} and Slingshot-11 interconnect~\cite{cug2022duncan}. The HPE Cray EX cabinet~\cite{hpe_ex} consists of 8 chasses with integrated switches and power distribution units. Rosetta is a 64-port ethernet switch implemented as a large monolithic (685 mm\textsuperscript{2}) ASIC fabricated in the TSMC 16 nm FinFet process and housed in a 62.5 mm package. All main switching logic utilizes an 850 MHz clock, which results in a typical power dissipation of 160~W and max of $\approx$ 300~W.

\begin{figure*}[ht]
    \centering
    \includegraphics[width=\linewidth]{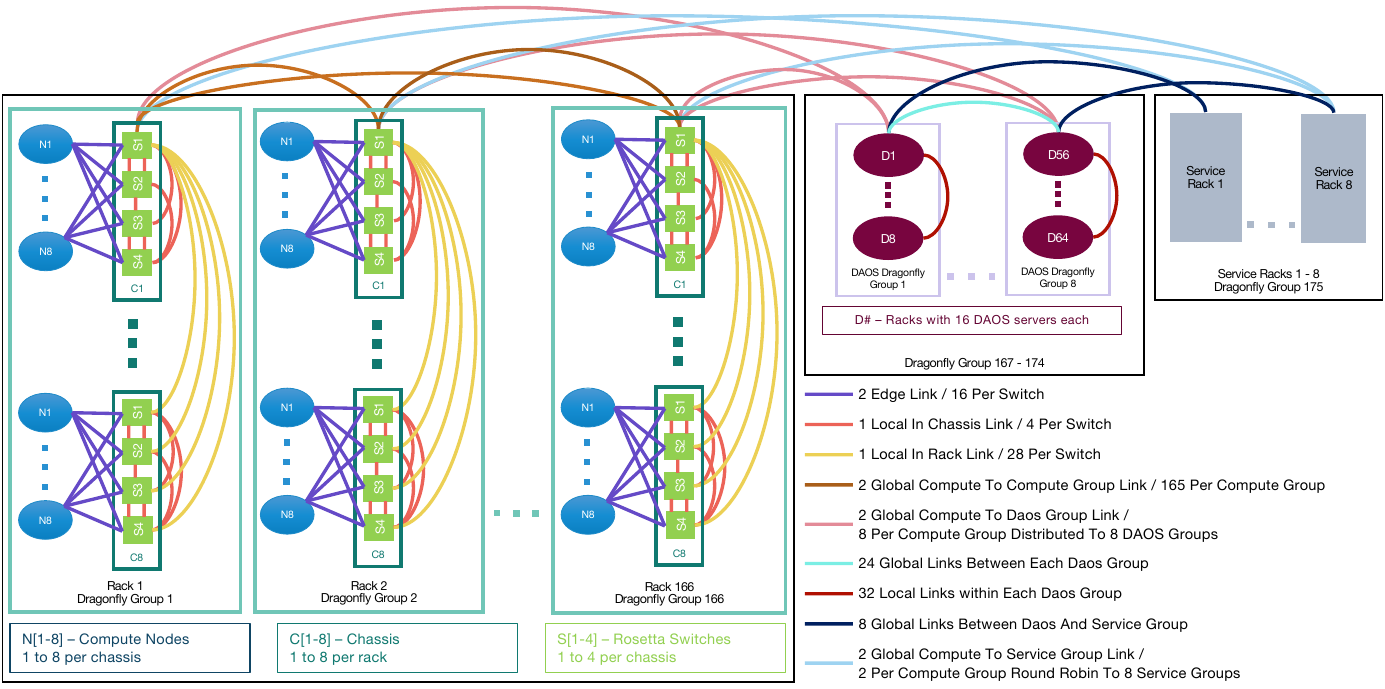}
    \caption{Aurora 1-D Dragonfly Topology.}
    \label{fig:aurora-network}
\end{figure*}
\subsubsection{1-D Dragonfly Topology} Aurora's network architecture follows a single dimension dragonfly topology that uses all-to-all local groups connected to each other through global links~\cite{4556717}. It consists of 175 total groups composed of 166 compute groups, 8 storage groups, and 1 service group as shown in figure~\ref{fig:aurora-network}. Each chassis consists of 4 switches connected to 8 compute nodes distributing the 4 ports per socket evenly among the switches, providing an injection bandwidth 2x 200~Gb link to each switch from the compute node. In the chassis consisting of 4 switches, 2 of the switches are connected with 2 links while the remaining switches are connected with a single link. In total 4 of the ports of the switches are dedicated for the local link within a chassis. Each switch in a chassis uses 28 ports to connect to every other switch in the rack forming an all-to-all connected group. The first switch in the rack is connected to the service group with two links. The next 8 switches are connected to the 8 DAOS I/O groups with two links. 10 ports in all switches and 10 additional ports in the 11 - 15 switches for a total of 330 links connect to all the 166 compute groups, providing 2 global links between each compute group.
Each compute group correspond to a single HPE Cray EX cabinet and  consists of 32 switches that are all-to-all connected using one network link between each pair of switches. Each storage and service group are spread across multiple River cabinets and contain 32 switches that are all-to-all connected using one network link between each pair of switches. All intra-group cables in the compute cabinets are electrical. Intra-group cables within the storage groups consist of a mix of electrical and optical.
The 166 compute groups contain 84,992 endpoints in the system that deliver 2.12 PB/s of injection bandwidth.
At the global level, it provides a total of 1.38 PB/s of global bandwidth and 0.69 PB/s of global bisection bandwidth between compute groups.



\subsection{Storage Subsystem}

\subsubsection{DAOS}
Aurora's primary storage system is based on the Distributed Asynchronous Object Store, DAOS~\cite{hennecke2020daos}.
DAOS is a new storage system that is open source and is specifically designed for high performance Flash and Storage Class Memory technologies.
The Aurora DAOS storage system~\cite{Latham2025DAOS} consists of 1024 Intel Coyote Pass dual-socket Ice Lake servers each with (16) 15.3~TB NVMe drives and (16) 512~GB Intel Optane PM200 DIMMs with (2) HPE SS200 NICs.
This hardware configuration enables 2048 DAOS engines with a theoretical peak performance of $\geq$31~TB/s and 400~MIOPs.
The Aurora DAOS system has been benchmarked with IO-500~\cite{IO500} and is currently first overall on the production list using 660 servers.
The total raw capacity of the DAOS is approximately 260~PB with no data protection.
Users can choose their level of data protection and a per-container basis, but assuming the all users choose the ALCF suggested 16 data + 2 parity (EC\_16P2GX) setting, that results in approximately 220~PB of total storage.



\subsubsection{Lustre}
Aurora is also connected to one of ALCF's Lustre storage systems, called "Flare".
This system provides an aggregate 100~PB of storage which provides additional capacity for Aurora as well as a simple method to share data between other ALCF resources.
Flare is a HPE E-1000 based storage system identical to another ALCF storage system, Eagle~\cite{eagle}, with 56 OSTs and 40 MDTs each.
Flare's peak performance is around 650~GB/s.
The system also features 100 gateway nodes that are part of the service group and provide connectivity to Lustre and support data movement capabilities to transfer data between Lustre and DAOS.


\section{Software and Programming Environment}\label{SoftwareAndProgrammingEnvironnment}
\label{AuroraProgrammingEnvironment}
The programming environment on Aurora is based on Intel's oneAPI \cite{oneapi} software ecosystem along with additional software components selected by ALCF from 3rd party vendors and open source components from projects such as the Exascale Computing Project. Aurora is the first large scale system to utilize Intel's oneAPI software suite, which offers a full featured environment for HPC application development providing compilers, libraries, tools, and AI learning frameworks. The focus of oneAPI is to provide a programming environment that facilitates development of applications that can run on a variety of hardware including Intel CPUs and GPUs and efficiently offload compute kernels to an accelerator in order to enhance compute performance.

The oneAPI compilers represent significant revisions of Intel's C/C++ and Fortran compilers, now invoked as icx/icpx/ifx instead of icc/icpc/ifort. The new C/C++ compilers are built on top of clang from the LLVM project, and support the SYCL programming model and OpenMP offload to accelerators. The Fortran compiler also supports OpenMP offload for accelerator programming. The oneAPI's compilers  leverage Intel’s Level Zero (\texttt{L0}) specification---a core component of oneAPI---as the primary runtime driver library, providing a low-level hardware interface to support cross-platform abstractions. In addition, Intel developed new versions of their math libraries, the Intel oneAPI Math Kernel Library (oneMKL)~\cite{MKL}. This library provides optimized Intel GPU implementations of dense and sparse BLAS and LAPACK, FFTs, random number generator, and other math library functions. It has APIs both in SYCL (C++) and OpenMP offload (Fortran, C, C++). The Unified Acceleration (UXL) Foundation defines the oneAPI specification, which includes APIs to drive high-performance libraries across domains such as mathematics, communication, data analytics, AI, and machine learning.

The MPI utilized on Aurora is an open source MPICH~\cite{doi:10.1177/10943420241311608}  implementation. It has been designed to leverage the hardware features of the Cassini NIC to achieve low latency and high throughput data transfers~\cite{mpich-sc17}. Some of the optimizations developed for Aurora include: high radix algorithms to decrease latency at large scale for collective operations~\cite{mpich-high-radix21}; efficient support for multi-threaded communication~\cite{mpich-threading}; GPU data transfers through Level Zero Inter Process Communication utilizing XeLinks; GPU Direct RDMA to avoid copies between CPU and GPU; GDRCopy support for small GPU data transfers; Yaksa~\cite{yaksa}, a GPU data copy engine to support data exchange of non-contiguous GPU buffers, and optimizations to leverage the multiple NICs in a single node. To support scalable launch of MPI jobs on thousands of nodes, the PMIx application programming interface~\cite{pmix} is used in combination with HPE Cray Parallel Application Launch~\cite{pals}. 



\subsection{Programming Models}\label{ProgrammingModels}

The Aurora system supports a number of programming models in order to enable a broad range of applications to run on the system. A key programming model on the Aurora system is SYCL, which is a programming model introduced in 2015 and standardized by the Khronos Group. SYCL provides abstractions using modern ISO C++ for an open-source, high-level, single-source programming standard designed to enable heterogeneous parallel computing. oneAPI offers a unified C++/Data Parallel C++ (DPC++) compiler, enabling both ISO C++ and SYCL source code to target Intel’s CPUs and GPUs on the Aurora platform. A key feature of oneAPI’s SYCL implementation is its provision of API extensions to enable new experimental features, typically focused on portability and performance, before they are considered  in the SYCL standard. 

In addition to SYCL, the oneAPI environment provides two other open standard programming models - OpenMP and OpenCL. Intel compilers support OpenMP 5+\cite{openmp}, including the most common OpenMP offload clauses to target GPUs, like `omp target map` for data transfer and `omp target [teams distribute parallel for]` for kernel launches. An OpenCL 3.0 implementation is provided and exposes a number of Khronos and Intel extensions allowing efficient implementation of compute run-times on top of OpenCL. 

While the above programming models provide a diverse set of open and standardized approaches to programming GPU systems, models such as NVIDIA's CUDA and AMD's Heterogeneous Interface for Portability (HIP)\cite{hip} programming model have been used widely within the HPC community. To support these applications an implementation of HIP is experimentally provided on Aurora by chipStar \cite{chipStar,chipStar_cl} - an Argonne-led open source project which implements the HIP API on top of the OpenCL and Level Zero runtimes and uses SPIR-V as a portable device code representation. The chipStar project also provides commonly used HIP libraries (e.g. hipBLAS) and an experimental CUDA implementation.

In addition to the above models, two DOE developed programming models, Kokkos and RAJA, are provided which have a focus on enabling portability across hardware platforms. Kokkos is a performance portability ecosystem developed as a modern, pure ISO C++ library that can target various backend platforms including a SYCL back-end developed for Aurora~\cite{kokkos_sycl}. RAJA provides portable parallel kernel execution as a C++ abstraction library and insulates application loop kernels from target architecture and programming model-specific implementation details. Architecture specific programming models, such as SYCL \cite{raja_sycl}, are utilized as back-ends in the RAJA library. Control over the target backend, the division of work, and loop transformations are exposed to the developer as execution policy template parameters. 


\subsection {Optimization and Debugging Tools}
Due to the tight coupling with the system hardware and low level software components, significant development was undertaken to make a number of tools from Intel and the wider HPC community available on Aurora. To provide debugging capabilities Intel's gdb-oneAPI has been extended to support Intel GPUs. Additionally, the parallel Linaro DDT debugger offers a graphical user interface (GUI) on top of gdb-oneAPI which provides a simple, easy-to-use method to debug applications. Additionally, the DDT debugger is available to support debugging applications at scale.  


A number of performance tools have been developed and extended to support performance analysis across the Aurora system and on PVC GPUs. Intel Application Performance Snapshot (APS) has been developed to provide an aggregated view of application performance for large scale MPI workloads. For more detailed analysis at small scale the Intel VTune profiler has been extended and optimized for multi-GPU analysis on Aurora compute nodes and performs various analyses such as GPU offload cost analysis, GPU hotspots analysis, source-level in-kernel profiling for instruction counts, and hardware-assisted stall sampling. In addition, Intel Advisor provides guidance on efficient GPU offloading and roofline performance analysis. Intel's unitrace tool~\cite{unittrace}  profiles layers in the software stack such as SYCL and plugins, oneCCL, and MPI for scale-up and scale-out applications on Intel GPUs. 


In addition to Intel tools, a broad community of HPC tools developers have collaborated with Argonne and Intel to add support for Intel GPUs into their tools. This includes HPCToolkit, TAU and PAPI. The THAPI (Tracing Heterogeneous APIs)~\cite{bekele2025thapi} is a portable tracing framework developed by Argonne for analysis of heterogeneous HPC applications and runtimes. It supports Intel oneAPI Level Zero as well as CUDA, OpenCL, HIP, OpenMP, and MPI and provides a holistic view of system and hardware behavior in the context of applications and runtimes -- enabling programming model centric debugging, power/energy optimization, and efficient resource management. 

To optimize power utilization the GEOPM (Global Extensible Open Power Manager)~\cite{geopm},  hierarchical power management framework was developed with Intel and supports Intel CPUs and GPUs on large scale systems. It features a service for low-level telemetry and control, and a runtime for profiling and dynamic tuning—such as power caps and frequency scaling—via agent-based plugins. 





\subsection{AI and Data Software Environment}
\label{AuroraAIData}

\subsubsection{AI frameworks} PyTorch and Tensorflow are natively and fully supported on Aurora's SPR CPUs and PVC GPUs (also referred to as XPU within oneAPI); these frameworks are optimized using libraries such as oneDNN, oneMKL, and the DPC++ compiler. While the AI and data software frameworks support GPU-as-a-device, wherein both stacks can be used as a single XPU device for kernel submission and thus providing higher memory capacity, the current libraries are highly optimized for stack-as-a-device, wherein each stack is a distinct XPU device.  The AI frameworks support just in time (JIT) compilation and graph mode to optimize performance on the XPU. 
A key enabler of distributed AI and ML at scale on Aurora is the \emph{oneAPI Collective Communications Library} (oneCCL)~\cite{oneCCL}. Built atop Xe-Links for intra-node communication and Slingshot for inter-node connectivity, oneCCL provides a highly optimized and flexible set of collective primitives, including broadcast, reduce, allreduce, allgather, and others—tailored for Intel's XPU architecture. Its integration with the major deep learning frameworks (e.g., PyTorch, TensorFlow, DistributedDataParallel (DDP), Horovod, DeepSpeed, etc.) enables large-scale training and inference workflows to efficiently exploit both on‑node and inter‑node parallelism. Here, oneCCL serves as the backend to route collective operations among the SPR CPU cores and PVC GPU stacks, transparently leveraging the most effective transport paths. This design is especially critical for Aurora’s multi-GPU-per-node configuration, where oneCCL's optimized ring, tree, and hybrid algorithms can dynamically adapt to diverse data sizes, topologies, and node counts.  Native PyTorch and TensorFlow inference is supported. Moreover, optimized inference  for language models is provided by frameworks such as OpenVINO~\cite{OpenVINO}, vLLM~\cite{vllm}, and SGLang~\cite{SGLang}.

The Hugging Face Transformer library ~\cite{HFT} and other significant ecosystems, such as TRL~\cite{TRL}, Accelerate~\cite{Accelerate} and TorchTune~\cite{TorchTune}, are supported to enable training, post-training and fine-tuning of large-language and foundation models. 
The OpenXLA compiler supports the XPU device via a PJRT plugin~\cite{openXLA} to enable JAX-based applications to run on Aurora.

\subsubsection{Python and Data Stack} On Aurora, users can access Intel's Python stack comprising of compilers and libraries for programming heterogeneous devices, namely the Data Parallel Extensions for Python (DPEP)~\cite{DPEP}. DPEP is composed of three main packages for programming on CPUs and GPUs: \textbf{1) dpnp} - Data Parallel Extensions for Numpy~\cite{dpnp} is a library that implements a subset of Numpy that can be executed on any data parallel device. The subset is a drop-in replacement of core Numpy functions and numerical data types, similar to CuPy for CUDA devices.
 \textbf{2) dpctl} - Data Parallel Control library~\cite{dpctl} provides utilities for device selection, allocation of data on devices, tensor data structure along with Python Array API Standard implementation, and support for creation of user-defined data-parallel extensions. \textbf{3) numba\_dpex} - Data Parallel Extensions for Numba~\cite{dpex} is an extension to Numba compiler for programming data-parallel devices similar to developing programs with Numba for CPU or CUDA devices.
The DPEP packages follow the compute-follows-data programming model, meaning that the offload target for a Python library call, or a hand-written kernel using numba-dpex, does not need to be specified directly when making the call. Instead, the offload target is inferred from the input arguments to the library call, meaning the user only needs to specify the offload target when creating the tensor/ndarray objects. The DPEP packages also provide interoperability with the AI frameworks through DLPack~\cite{DLPack}, with support for zero-copy array creation on XPU devices. 

The oneAPI Data Analytics Library (oneDAL)~\cite{onedal} is a powerful machine learning (ML) library that helps to accelerate big data analysis. The library implements many classical ML algorithms, with support for both single-GPU and multi-GPU, multi-node workloads through oneCCL. OneDAL drives the scikit-learn accelerator engine infrastructure and provides support  on the XPU device and enables users to execute scikit-learn based applications, commonly used for data analytics, on Aurora.   To support big-data processing, Apache Spark is an avenue for users on Aurora. This has been integrated with DAOS to support HDFS. Spark MLlib has been optimized using oneDAL and leverages oneCCL to scale on Aurora.

\subsubsection{Scaling Python environments} Frameworks are made available to the users through a conda environment (with pip installed packages) provided in the compute image of an Aurora node. However, this is a challenge when the compute image doesn't contain all the necessary packages needed by an application or newer versions of these are needed - the typical use-case in production. At the scale of Aurora, application startup becomes a bottleneck due to thousands of processes simultaneously loading dynamic shared libraries. For instance, even a simple `import torch' across nodes can overwhelm the parallel file system (PFS) metadata servers, causing a substantial system-wide idle time. Copper~\cite{Copper, Copper_SC24} is a tool designed to solve this problem via a cooperative caching layer that significantly reduces the initialization time - often to under a minute - saving time and system resources. Copper has been deployed in production and reduces load times of python environments at scale by up to 90\%, with no changes to the application.

\subsubsection{Workflows and Visualization} On Aurora, science applications can benefit from a number of workflow tools to launch and manage ensembles (e.g., PBS job arrays), high-throughput campaigns (e.g., Balsam, Parsl, libEnsemble), and for coupling simulation and AI workloads (e.g., SmartSim and ADIOS2). To provide an intuitive experience and lower the barrier to entry, Jupyter notebooks are supported.  Aurora’s GPUs include hardware ray tracing cores and make them well-suited for advanced rendering. The OSPRay ray tracing framework~\cite{7539599} has been ported to leverage the GPUs. Additionally, VTK-m~\cite{moreland2024visualization} has been optimized to support visualization algorithms on the GPUs. Together, these have enabled scientific visualization packages such as ParaView, Visit, and Ascent to run at scale on Aurora.




\section{Early Applications}\label{EarlyApplications}

Two main application-readiness programs drove the efforts to port and optimize scientific applications for Aurora: ALCF's Aurora Early Science Program (ESP)~\cite{ESP} and the U.S. Department of Energy's Exascale Computing Project (ECP)~\cite{ECP}. 
The Argonne-Intel Center of Excellence (COE) worked with ALCF staff to support projects across these two programs. The COE organized 30 training events for ESP project teams, including intensive hands-on hackathons. ALCF tracked progress in porting, tuning, and single-GPU performance via an Aurora Applications Working Group that met bi-weekly and surveyed these applications projects quarterly.

\subsection{Early Science Program, ECP, and Year One Production Projects}\label{EarlyScienceProgramECPandYear1ProductionProjects}

The Early Science Program supported 19 projects and accounted for ALCF's mission to support the three pillars (Section~\ref{Introduction}) as peers~\cite{Kim_2018,Uintah,XGC1,XGC2,XGC3,MILC,Chroma,Grid,QUDA,HABIB201649,PHASTA1,PHASTA2,NAMD,nwchemex,BYLASKA2024518,FastCaloSim,MadGraph,cfdml,HARVEY,LQCDML,connectomics,BerkeleyGW,SST}. The projects had six years of access to pre-PVC testbed hardware and pre-release oneAPI software, which they helped debug and harden for production. This, along with producing strong candidates for early production projects, were key goals of the ESP. Each project proposed an exascale science campaign and gained access to run it in January 2025---the same time that the first year Innovative and Novel Computational Impact on Theory and Experiment (INCITE) and Advanced Scientific Computing Research (ASCR) Leadership Computing Challenge (ALCC) projects started.

Of the 21 ECP Application Development (AD) projects, 15 partnered closely with ALCF to port and tune applications for Aurora. Several ECP Software Technology (ST) projects, spanning Data and Visualization, Mathematical Libraries, Programming Models and Runtimes, Development Tools, and Software Ecosystem and Delivery, worked closely with ALCF to port and optimize for Aurora. These project teams were supported with staff points of contact. 



With the transition of Aurora to Production and general availability on 27 January 2025, ALCF has 34 production INCITE projects. For the first half-year, ALCF has 13 production ALCC projects. ESP PIs, project team members, and applications are represented in 15 of the first-year INCITE projects and 3 of the first-year ALCC projects on Aurora. ECP PIs, team members, and applications are represented in 14 first-year projects.

\subsection{Portable Programming Choices for Aurora}

\subsubsection{Portability}
All of the ESP, ECP, INCITE, and ALCC projects targeted performance portability across the pre-exascale, and future post-exascale supercomputing systems for science in the U.S. This includes the major vendors' CPUs and GPUs (Intel, AMD, NVIDIA, and Arm). Separate implementations for different platforms are impractical for the sophisticated applications targeting ALCF. Performance portability requires using and/or developing a \textit{portability layer} to program the GPUs.

\subsubsection{Compiled-Code Portability Layers}
Section~\ref{ProgrammingModels} covers GPU programming options for compiled-language codes typical of simulation applications. OpenMP, Kokkos, and RAJA are portability layers that pass through to the optimized Intel GPU runtime, directly or via SYCL. While SYCL could stand as a portability layer, none of the Aurora science projects have adopted it that way---though there has been work comparing application performance using SYCL across different GPU architectures~\cite{nwchemex_sycl}. Some projects implemented their own portability layer; examples include (1) the OCCA macro kit used by the NekRS application, the gtensor library for the GEM application, and the TAMM~\cite{nwchemex_tamm} library for NWChemEx. Mostly, the portability layers feed through to SYCL on PVC, and to HIP/CUDA on AMD/NVIDIA GPUs.

\subsubsection{Data and Learning Portability Layers}
%
The data-intensive and AI/ML programming community, scientific and nonscientific, has broadly adopted the ubiquitous Python frameworks discussed in~\ref{AuroraAIData}. Driven by the explosively-growing AI market, major hardware vendors have developed optimized implementations of the high-level frameworks that perform well on their CPUs and GPUs. Intel is no exception, and has developed the software discussed in \ref{AuroraAIData}.

\subsubsection{GPU Programming Approaches Chosen by ESP, ECP, and Year-One Aurora Projects}


Many projects include either multiple applications for different subtasks in science campaigns or workflows with multiple components. In the ESP projects, some components are compiled simulation applications and others are AI models trained and/or used for inference. Each application or workflow component may use a different programming approach, so a single project may use multiple programming approaches.

\begin{figure}[htbp]
    \centering
    \includegraphics[width=\linewidth]{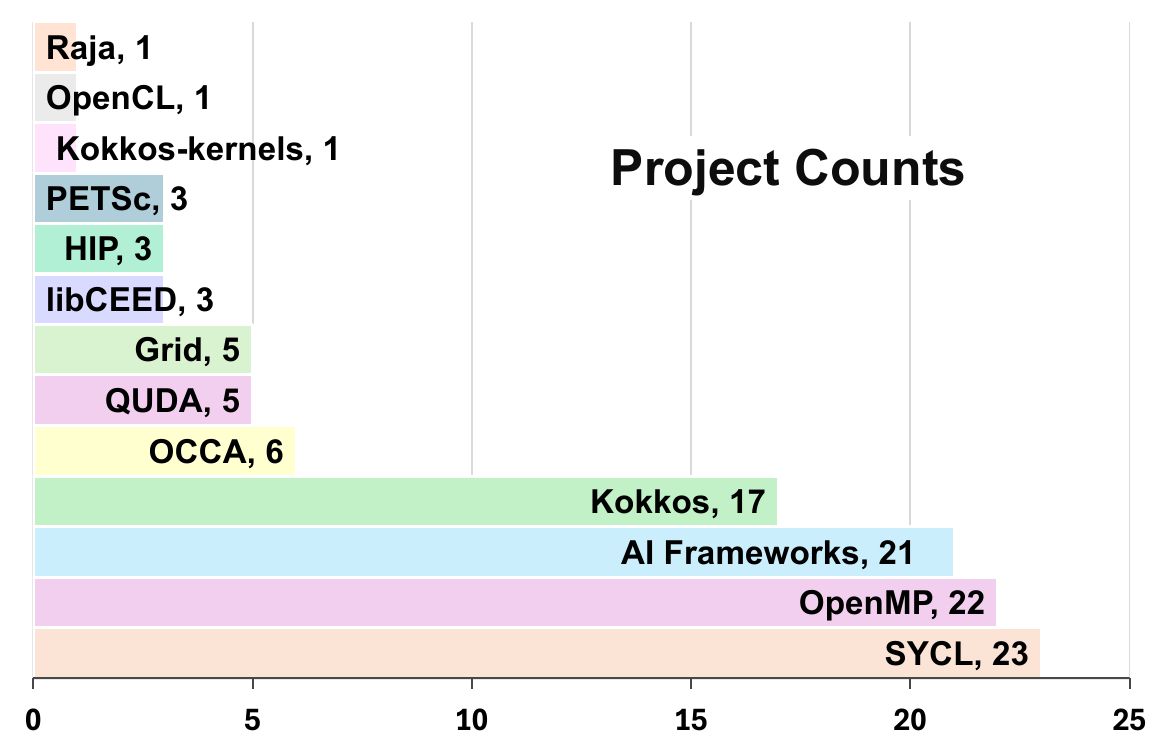}
    \caption{Breakdown of programming approach choices across all the year-one Aurora science projects in 2025.}
    \label{fig:ProgrammingApproachChoicesPieChart}
\end{figure}

Figure~\ref{fig:ProgrammingApproachChoicesPieChart} summarizes data across all of the year-one Aurora science projects in the ESP, INCITE, and ALCC programs, and the 15 ECP AD projects discussed in \ref{EarlyScienceProgramECPandYear1ProductionProjects}. Where an application or workflow component is common to multiple projects, its programming approach is counted once for each project. SYCL usage is primarily via its use in portability layers and libraries leveraged by applications rather than an application directly using SYCL. Note that all but one of the Fortran applications chose OpenMP. One Fortran application (Fun3D) uses macros to call SYCL kernels, and a second application (BerkeleyGW) calls SYCL functions for a subset of kernels and uses OpenMP for the rest. Two C++ applications (OpenMC and QMCPACK) chose to use OpenMP (one uses SYCL indirectly via oneMKL library calls). See \ref{secA2} for the project-by-project breakdown.


\subsection{Performance of Early Applications}

Because of the relative timings of ECP's end in 2023 and Aurora's first availability to applications teams late in that year, only a limited number of ECP AD projects measured their performance Figure of Merits (FOMs) on Aurora. Those that did yielded strong results, as shown in Table~\ref{tab:AuroraECPFOMs}.


\begin{table}[htbp]
    \centering
    \caption{Measured ECP FOM ratios on Aurora for EXAALT (LAMPS molecular dynamics run with SNAP potential), ExaSMR (OpenMC Monte Carlo transport case), and ExaSky (gravity+hydrodynamics cosmology simulation). }
    \begin{tabular}{lrc}
        \toprule
        \textbf{Project} & \textbf{Nodes} & \textbf{FOM Ratio} \\
        \midrule
        EXAALT & 1024 & 89x \\
        ExaSMR & 512 & 84x \\
        ExaSky & 4096 & 277x \\
        \bottomrule
    \end{tabular}
    \label{tab:AuroraECPFOMs}
\end{table}

In contrast with the performance FOMs targeted by ECP, the goal of the ESP was to demonstrate computational readiness at the level expected by an INCITE proposal. The ESP projects were reviewed internally and externally against criteria used by the LCF staff in Technical Assessment (TA) of INCITE proposals (formerly known as Computational Readiness review). Our review determined that 16 of the 19 projects achieved a passing grade for INCITE TA. In Fall 2024, all 16 demonstrated scaling up to at least 2000 Aurora nodes, which was the 20\% threshold  required for capability computing in the 2025 INCITE TA.

The Applications Working Group asked all the ESP and ECP projects to define, for their chief GPU-offloaded applications or workflow components, a single-GPU performance FOM for which higher is better. 
Certain relative levels of performance on PVC were expected, based on hardware specifications for PVC and peer contemporary GPUs: 
comparable performance with AMD MI250X, and higher performance than NVIDIA A100. 
From tracking single-GPU FOM measurements across the early applications projects over 3 years, we believe PVC met such expectations.

\begin{figure}[htbp]
    \centering
    \includegraphics[width=0.8\linewidth]{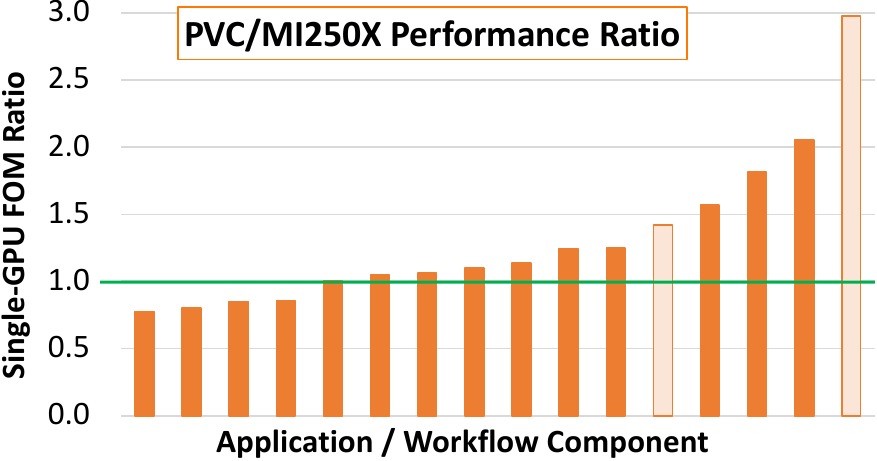}
    \includegraphics[width=0.8\linewidth]{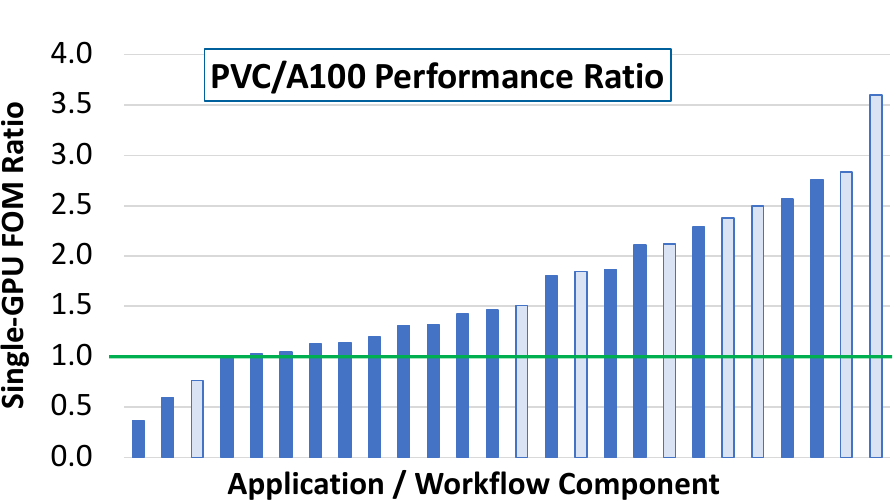}
    \caption{Ratios of single-GPU performance FOM 
    across subsets of ESP and ECP applications. Each bar is one application.}
    \label{fig:SingleGPUFOM}
\end{figure}

Figure~\ref{fig:SingleGPUFOM} shows a performance snapshot from early 2024 on subsets of the applications spanning the three pillars compared to other GPUs. Light bars were measured on one PVC stack and multiplied by 2; dark bars were measured on both PVC stacks; MI250X was measured on both Graphics Compute Dies (GCD).


\section{Performance Evaluation}\label{PerformanceEvaluation}
Performance is evaluated from node level to scalable benchmarks and applications based on the system architecture and design. 
\subsection{Node-Level Performance}

This section highlights the compute and memory performance of the individual components within an Aurora node.
%
%
Table~\ref{tab:gemm_stream_gpu_cpu} presents the GEMM performance of a single Intel® Xeon® Max SPR CPU and a single Intel® Data Center GPU Max across a variety of data types. The results are the median of the measured peak performances across a single-node sweep of all the GPUs and CPUs on Aurora available at the time (see ~\cite{gemm_benchmarking}) except for FP64 on GPU, which is from a 512-node run.  
The PVC GPU supports all the numerical precisions available on the SPR CPU, with the addition of TensorFloat-32 (TF32). 
The table also shows the measured memory bandwidth using the STREAM Triad benchmark on a single socket, for both high-bandwidth memory (HBM) and traditional DDR memory, and on a single Intel® PVC GPU for HBM.

\begin{table}[htbp]
\centering
\caption{Aurora median of measured peak performance on an Intel® PVC GPU and on an Intel® Xeon® Max CPU.}
\begin{tabular}{lcc}
\hline
\textbf{GEMM-Data Type}    & \multicolumn{2}{l}{\textbf{Performance (TF/s)}}           \\
                      & \textbf{CPU}           & \multicolumn{1}{c}{\textbf{GPU}} \\ \hline
FP64 (DGEMM)          & 2.9                    & 29.2                                \\
FP32 (SGEMM)          & 5.7                    & 44.0                               \\
TF32 (SGEMM-TF32)     & N/A                    & 212.8                              \\
BF16 (HGEMM-BF16)     & 22.9                   & 420.3                              \\
FP16 (HGEMM-FP16)     & 10.9                   & 421.6                              \\
INT8 (IGEMM)          & 65.5                   & 854.3                              \\
                      &                        &                                  \\ \hline
\textbf{STREAM triad} & \multicolumn{2}{l}{\textbf{Bandwidth (TB/s)}}             \\
                      & \textbf{CPU}           & \multicolumn{1}{c}{\textbf{GPU}} \\ \hline
HBM2e                 & 0.63                   & 2.1                              \\
DDR5                  & 0.24                   & N/A                               \\ \hline
\end{tabular}
\label{tab:gemm_stream_gpu_cpu}
\end{table}

\subsection{Scalable Benchmark Performance}
Standard HPC and AI benchmarks, as shown in Table~\ref{tab:Scalable_Performance_Benchmarks}, were run on Aurora to evaluate its performance at scale. These benchmarks include High-Performance Linpack (HPL) for the TOP500 list~\cite{TOP500}, which measures peak double-precision floating-point performance; HPL-MxP~\cite{HPLMxP} for mixed- and lower-precision floating-point operations typically used in AI workloads; HPCG~\cite{HPCG} for assessing computational performance and data access patterns; Graph500~\cite{GRAPH500}~for large-scale graph analytics; and IO500~\cite{IO500} for user experience and IO performance. Collectively, these benchmarks stress multiple aspects of Aurora's architecture, including computation, memory hierarchy and data movement, and network performance across interconnects.


\begin{table}[htb]
\centering
\caption{Scalable Performance Benchmarks on Aurora.}
\begin{tabular}{lllc}
\toprule
\textbf{Benchmark} & \textbf{Performance} & \textbf{Rank} & \textbf{\# of Nodes}  \\
\midrule
HPL     & 1.012 EF/s           & \#3 at SC24\   & 9234   \\
\midrule
HPL MxP & 11.64 EF/s           & \#1 at SC24   & 9500     \\
\midrule
HPCG  & 5.613 PF/s        & \#3 at ISC24   & 4096     \\
\midrule
Graph500  & 24,250 GTEPS        & \#6 at ISC24   & 4096     \\
Graph500  & 69,373 GTEPS           &                & 8192     \\
\midrule
IO500     & 32,165.90            & \#1 at SC23      & 300\\
\bottomrule
\end{tabular}
\label{tab:Scalable_Performance_Benchmarks}
\end{table}

\subsubsection{HPL}
\begin{figure}[htb]
    \centering
     \includegraphics[width=0.94\linewidth]{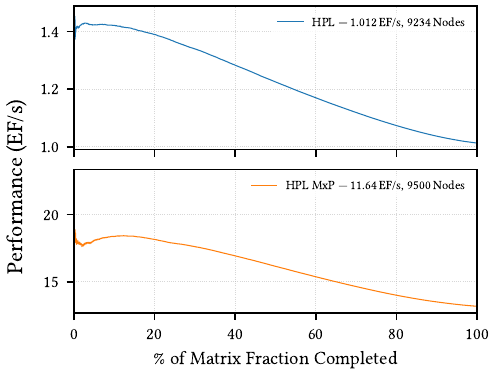}
     \caption{Performance of HPL at 9,234 nodes and HPL MxP at 9,500 nodes.}
     \label{fig:HPL_HPL_MXP_Performance}
 \end{figure}

Aurora achieved a performance of 1.012 EF/s using 9,234 nodes (\(\sim87\%\) of the full system), ranking third on the TOP500 list at the time of submission for SC24. The MPI process grid was configured with \( P = 162 \) and \( Q = 342 \), where \( P \) and \( Q \) represent the number of MPI processes in each dimension. This configuration resulted in a scaling efficiency of 78.84\%, defined as the ratio of achieved performance per node to the expected peak performance per node. As shown in Figure~\ref{fig:HPL_HPL_MXP_Performance}, performance remained relatively smooth across all phases of computation, from dense lower-upper (LU) factorization to the iterative refinement (IR) phase, and finally to the computation of the score. In the initial phase, there was slight performance degradation, highlighting an opportunity for further optimization to improve scaling efficiency.


\subsubsection{HPL-MxP}
Aurora achieved a performance of 11.64 EF/s using 9,500 nodes (\(\sim89\%\) of the full system), ranking first on the HPL-MxP list~\cite{HPLMxP} at the time of submission for SC24. The LU factorization phase utilizes mixed-precision data types, FP16 and FP32, while the iterative refinement (IR) phase uses FP64. As shown in Figure~\ref{fig:HPL_HPL_MXP_Performance}, performance scales uniformly across the various phases of computation and communication. However, there is potential for further performance improvement, as slight degradation was observed in the initial and final phases. Optimizing for compute and communication resources, reducing latencies during broadcast and swap operations, could enhance performance in these phases.

\subsubsection{MPICH Performance}
MPICH has been optimized to deliver low latency and high throughput. Table~\ref{tab:mpich-performance} shows performance of some MPICH microbenchmarks with CPU and GPU buffers. 


\begin{table}[ht]
\centering
\caption{MPICH Microbenchmarks.}
\begin{tabular}{lclc}
\toprule
\textbf{Benchmark} &  \textbf{Location} & \textbf{Message Size} &  \textbf{Result}  \\
\midrule
                  & CPU    & 0 B           &  1.9~$\mu s$   \\
PingPong          & CPU    & 4 KiB         &  3.3~$\mu s$   \\
Latency           & CPU    & 64 KiB        &  5.9~$\mu s$   \\
\cmidrule{2-4} 
(0 hops)          & GPU    & 4 KiB         &  4.0~$\mu s$   \\
                  & GPU    & 64 KiB        &  6.6~$\mu s$  \\

\midrule
Bandwidth (1 NIC)  & CPU  &  512 KiB      &    23.5  GB/s\\
Bandwidth (4 NICs) & CPU  &  512 KiB      &    94.7  GB/s\\
\cmidrule(lr){2-4}
Bandwidth (1 NIC)  & GPU  &  512 KiB      &    23.0  GB/s\\
Bandwidth (4 NICs) & GPU  &  512 KiB      &    35.9  GB/s\\
\midrule
Allreduce (8192 nodes)  & CPU  &  8 B     &    53.8~$\mu s$ \\
\cmidrule(lr){2-4}
Allreduce (8192 nodes)  & GPU  &  8 B     &    60.5~$\mu s$ \\
\bottomrule
\end{tabular}
\label{tab:mpich-performance}
\end{table}

\subsubsection{oneCCL Performance}
\begin{figure}[htbp]
    \centering
    \includegraphics[width=0.90\linewidth]{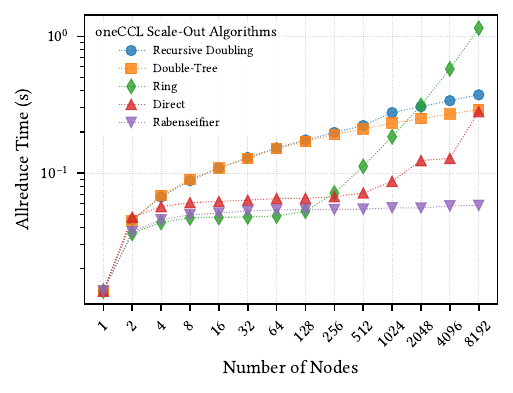}
    \caption{oneCCL allreduce time for 1 GB, 12 ranks/node.}
    \label{fig:oneCCL_bench}
\end{figure}

The oneCCL collectives are similar to MPI but add a stream parameter for tighter synchronization between compute kernels and communication~\cite{petsc-paper}. They are implemented in two phases: \emph{scale-up} (for GPUs in the same node) and \emph{scale-out} (for GPUs across nodes). Scale-up algorithms are tuned for the PVC all-to-all topology, while scale-out supports several algorithms, including \emph{ring}, \emph{recursive doubling}, and \emph{Rabenseifner} ~\cite{thakur_collectives}. For data transport, oneCCL uses either MPI or OFI; with MPI, it may call MPI directly for scale-out (shown as \emph{Direct} in Figure~\ref{fig:oneCCL_bench}). All performance data in Figure~\ref{fig:oneCCL_bench} were collected using the oneCCL~2021.14 benchmark with MPI transport, reporting the minimum time over 100 iterations after 10 warm-up steps. For \emph{Rabenseifner}, which primarily targets powers-of-two node counts, the measured time remains flat as the number of nodes increases. This is expected as the algorithm is bandwidth constrained for large message sizes. In contrast, the time for \emph{ring} increases since the overhead incurred by passing messages scales linearly with node count, even if total message size stays the same.



\subsection{Application Benchmark Performance}

Performance on Aurora was evaluated using a diverse set of reference applications spanning three key computational pillars: simulation, data analytics, and machine learning. The simulation workloads are derived from the CORAL application benchmarks and, with minimal modifications, run successfully on Aurora. From a performance perspective, the primary objective is to achieve a $50\times$ or greater improvement in the figure of merit (FOM) relative to a reference system. This performance target corresponds to a transition from a 20-petaflop system to an exascale-class machine. The reference systems include Sequoia and Titan, each normalized to 20 PFLOPS. Table~\ref{tab:Applications_Performance} summarizes the applications, their performance gains relative to the reference systems, and the number of nodes.


\begin{table}[ht]
\centering
\caption{Reference application performance summary.}
\begin{tabular}{llcc}
\toprule
\textbf{Pillar} & \textbf{Benchmark} & \textbf{FOM Ratio} & \textbf{\# of Nodes}  \\
\midrule
Simulation     & HACC               &  54.4   &  9984  \\
               & LAMMPS             &  54.5   &  9216  \\
               & QMCPACK            &  60.2   & 10240  \\
               & Nekbone            &  22.2   &  8128  \\
\midrule
Data Analytics & TomoPy             &  56.7   &  8192  \\
               & ThunderSVM         &  81.6   &  8192  \\
               & FCMA               &  75.6   &  8192  \\
\midrule
Machine Learning & ResNet-50        &  42.2   &  9918  \\
               & Candle Pilot Uno   & 130.1   & 10240  \\
               & CosmicTagger       & 124.6   & 10240  \\
               & BERT               &  70.1   & 10240  \\
\bottomrule
\end{tabular}
\label{tab:Applications_Performance}
\end{table}

These applications utilize a variety of programming models and software stacks, collectively exercising the system across key aspects such as compute parallelism, data movement, and collective communication. For example, \textbf{HACC} features multiple memory-intensive phases, such as tree building, which benefit from the high-bandwidth memory (HBM) on Intel® Xeon® Max CPUs. Performance optimizations include bundling small data transfers over PCIe to reduce communication overhead.

For \textbf{Nekbone}, performance is primarily constrained by GPU memory bandwidth, limiting scalability compared to FLOP-bound computations. In the case of \textbf{CANDLE Pilot Uno}, irregular GEMM patterns and an I/O-intensive pipeline reduce single-instance efficiency. Aurora addresses these limitations by running 48 concurrent instances per node (4 per PVC stack), with each instance processing 23,290.5 samples per second. This throughput-oriented configuration is well suited for hyperparameter optimization, enabling many independent runs with distinct parameter sets.

\textbf{ResNet-50} is trained on the ImageNet dataset using PyTorch with Horovod. Each training instance spans 2,048 stacks (171 nodes) and processes 16 images per stack, resulting in a global batch size of 32,768. Aurora’s architecture enables high-performance distributed training through several key features: high single-stack performance from Intel® Data Center GPU Max (PVC), fast MDFI links between paired PVC stacks, high-bandwidth Xe-Links interconnecting six PVCs per node for efficient collective operations, eight NICs per node for inter-node communication, and high-capacity, high-bandwidth DAOS storage for fast access to training data.


\section{Challenges} \label{Challenges}

At very high scale, supercomputers face critical challenges in maintaining stability, scalability, and reliability — as increasing components' count and their complexities amplify the risk of faults, communication bottlenecks, and uneven workload distribution. 

Failure rates increase non-linearly, making proactive fault detection and recovery mechanisms essential to maintaining overall system reliability and availability. On Aurora, failure rates align with those observed in recent large-scale AI training infrastructures~\cite{cui25arxiv, meta24arxiv, kokolis25arxiv}. 
The main compute components on Aurora have been specifically designed with an enhanced level of Resilience, Availability, and Scalability (RAS) features to proactively address the unique demands of such a large-scale deployment and an automated failure management system was developed to monitor RAS events and conducts in-field repairs (IFR) autonomously ~\cite{bgerofi25MW}.

The system utilizes a centralized meta-database that aggregates data from various sources, including real-time component inventory, software/firmware versioning, failure events categorized by an automated system, maintenance operations, resource manager state transitions (e.g., compute nodes' queue assignments in the batch scheduler), and job submission data. Automation is driven by fine-grained multi-strike policies based on statistical properties of failure events, orchestrating diagnostics and IFR tools, interacting with the cluster resource manager and batch scheduler, and managing maintenance tickets for component replacements. Unreported errors, such as silent data corruption (SDC), are mitigated by screening with bit-wise reproducible applications and tests during idle- or maintenance-periods. A subset of theses tests is randomly chosen and run before each compute job.


\section{Conclusion}\label{Conclusion}

This paper provided an overview of the hardware, computing \& storage, and software architectures of Aurora, Argonne’s first exascale supercomputer, designed to support scientific advancements across 3 key areas: traditional HPC, data, and AI. 
The Aurora Early Science Program was discussed, highlighting the range of applications that have been successfully ported and optimized to run on this advanced platform.
Additionally, results from standard benchmarks were shared, illustrating Aurora’s capability to meet the performance metrics it was designed for, reinforcing its capabilities 
to enable advancing research and accelerating scientific breakthroughs in the era of exascale computing. Aurora was opened to general users and production science runs started in the 1st quarter of 2025.

\begin{appendices}

\section{Details about Early Applications}\label{secA}

\subsection{ESP and ECP AD Projects}\label{secA1}

These are the ESP projects and some of their key PVC-offloading applications and workflow components:

\begin{enumerate}[leftmargin=2em]

  \item{\it \href{https://alcf.anl.gov/science/projects/quantum-monte-carlo-foundation-models-exascale-driven-electronic-structure}{QMCPACK} (pillar:Simulation)(PI:Benali)}: QMCPACK \cite{Kim_2018}

  \item{\it \href{https://alcf.anl.gov/science/projects/design-and-evaluation-high-efficiency-boilers-energy-production-using-hierarchical}{Uintah} (pillar:Simulation)(PI:Berzins)}: Uintah \cite{Uintah}

  \item{\it \href{https://alcf.anl.gov/science/projects/high-fidelity-simulation-fusion-reactor-boundary-plasmas}{XGC} (pillar:Simulation)(PI:Chang)}: XGC \cite{XGC1,XGC2,XGC3}

  \item{\it \href{https://alcf.anl.gov/science/projects/lattice-quantum-chromodynamics-calculations-particle-and-nuclear-physics}{LatticeQCD} (pillar:Simulation)(PI:Christ)}: MILC \cite{MILC}, Chroma \cite{Chroma}, Grid \cite{Grid}, QUDA \cite{QUDA}, QDP++, QDP-JIT

  \item{\it \href{https://alcf.anl.gov/science/projects/extreme-scale-cosmological-hydrodynamics}{HACC} (pillar:Simulation)(PI:Heitmann)}: CRK-HACC \cite{HABIB201649}

  \item{\it \href{https://alcf.anl.gov/science/projects/metascalable-layered-materials-genome}{NAQMC\_RMD} (pillar:Simulation)(PI:Nakano)}: DCMESH, RXMD-NN

  \item{\it \href{https://alcf.anl.gov/science/projects/extreme-scale-unstructured-adaptive-cfd}{PHASTA} (pillar:Simulation)(PI:Jansen)}: PHASTA \cite{PHASTA1,PHASTA2}

  \item{\it \href{https://alcf.anl.gov/science/projects/free-energy-landscapes-membrane-transport-proteins-0}{NAMD} (pillar:Simulation)(PI:Roux)}: NAMD \cite{NAMD}

  \item{\it \href{https://alcf.anl.gov/science/projects/nwchemex-tackling-chemical-materials-biochemical-challenges-exascale-era}{NWChemEx} (pillar:Simulation)(PI:Windus)}: NWChemEx \cite{nwchemex}

  \item{\it \href{https://alcf.anl.gov/science/projects/exascale-computational-catalysis}{Catalysis} (pillar:Data)(PI:Bross)}: PWDFT \cite{BYLASKA2024518}\textbf{}

  \item{\it \href{https://alcf.anl.gov/science/projects/dark-sky-mining}{DarkSkyML} (pillar:Data)(PI:Habib)}: DarkSkyML, CRK-HACC

  \item{\it \href{https://alcf.anl.gov/science/projects/simulating-and-learning-atlas-detector-exascale}{ATLAS} (pillar:Data)(PI:Hopkins)}: FastCaloSim \cite{FastCaloSim}, MadGraph \cite{MadGraph}

  \item{\it \href{https://alcf.anl.gov/science/projects/data-analytics-and-machine-learning-exascale-computational-fluid-dynamics}{CFDML} (pillar:Data)(PI:Jansen)}: subgrid stress model \cite{cfdml}, PHASTA \cite{PHASTA1,PHASTA2}

  \item{\it \href{https://alcf.anl.gov/science/projects/extreme-scale-situ-visualization-and-analysis-fluid-structure-interaction}{Multiphysics} (pillar:Data)(PI:Randles)}: HARVEY \cite{HARVEY}

  \item{\it \href{https://alcf.anl.gov/science/projects/machine-learning-lattice-quantum-chromodynamics}{LqcdML} (pillar:Learning)(PI:Detmold)}: flow model \cite{LQCDML}

  \item{\it \href{https://alcf.anl.gov/science/projects/enabling-connectomics-exascale-facilitate-discoveries-neuroscience}{Connectomics} (pillar:Learning)(PI:Ferrier)}: FFN \cite{connectomics}

  \item{\it \href{https://alcf.anl.gov/science/projects/many-body-perturbation-theory-meets-machine-learning-discover-singlet-fission}{MatML} (pillar:Learning)(PI:Marom)}: BerkeleyGW \cite{BerkeleyGW}

  \item{\it \href{https://alcf.anl.gov/science/projects/virtual-drug-response-prediction}{VirtualDrugDesign} (pillar:Learning)(PI:Stevens)}: SST \cite{SST}

  \item{\it \href{https://alcf.anl.gov/science/projects/accelerated-deep-learning-discovery-fusion-energy-science}{FusionDL} (pillar:Learning)(PI:Tang)}: FusionDL \cite{FusionDL}

\end{enumerate}

\noindent The 15 AD projects that partnered with Argonne, and the relevant applications, were:

\begin{enumerate}[leftmargin=2em]
    \item {\it CANDLE:} Uno
    \item {\it EQSim:} SW4
    \item {\it E3SM:} E3SM-MMF
    \item {\it EXAALT:} LATTE and LAMMPS
    \item {\it ExaFEL:} SpiniFEL and cctbx
    \item {\it ExaSky:} Nyx and HACC
    \item {\it ExaSMR:} OpenMC and NekRS
    \item {\it ExaStar:} Flash-X/Thornado
    \item {\it ExaWind:} Nalu-Wind (runs on CPU), AMR-Wind, and OpenFAST (runs on CPU)
    \item {\it GAMESS:} GAMESS
    \item {\it LatticeQCD:} Grid, QUDA, Chroma, and MILC
    \item {\it MFIX-Exa:} MFIX-Exa
    \item {\it NWChemEx:} NWChemEx
    \item {\it QMCPACK:} QMCPACK
    \item {\it WDMApp:} XGC, GENE, GEM
\end{enumerate}

\subsection{Programming Model Choices}\label{secA2}

Tables \ref{tab:ESPProjectProgrammingApproachChoices}--\ref{tab:ECPProjectProgrammingApproachChoices} show the Aurora PVC programming approach choices for every relevant application or workflow component in every year--one Aurora science project, as well as the ECP AD projects that targeted Aurora specifically. \textit{TBD} indicates components under development for Aurora.

\begin{table}[htbp!]
\label{tab:ESPProjectProgrammingApproachChoices}
\caption {Aurora programming choice approaches for ESP projects.}
\begin{tabular}{|l|p{0.25\textwidth}|>{\RaggedRight}p{0.55\textwidth}|}
\hline
\textbf{Program} & \textbf{Project} & \textbf{Approaches} \\ \hline
ESP & \href{https://alcf.anl.gov/science/projects/quantum-monte-carlo-foundation-models-exascale-driven-electronic-structure}{QMCPACK} & OpenMP \\ \hline
ESP & \href{https://alcf.anl.gov/science/projects/design-and-evaluation-high-efficiency-boilers-energy-production-using-hierarchical}{Uintah} & Kokkos \\ \hline
ESP & \href{https://alcf.anl.gov/science/projects/high-fidelity-simulation-fusion-reactor-boundary-plasmas}{XGC} & Kokkos \\ \hline
ESP & \href{https://alcf.anl.gov/science/projects/lattice-quantum-chromodynamics-calculations-particle-and-nuclear-physics}{LatticeQCD} & QUDA, Grid \\ \hline
ESP & \href{https://alcf.anl.gov/science/projects/extreme-scale-cosmological-hydrodynamics}{HACC} & SYCL \\ \hline
ESP & \href{https://alcf.anl.gov/science/projects/metascalable-layered-materials-genome}{NAQMC\_RMD} & OpenMP, AI Frameworks \\ \hline
ESP & \href{https://alcf.anl.gov/science/projects/extreme-scale-unstructured-adaptive-cfd}{PHASTA} & libCEED, PETSc \\ \hline
ESP & \href{https://alcf.anl.gov/science/projects/free-energy-landscapes-membrane-transport-proteins-0}{NAMD} & SYCL \\ \hline
ESP & \href{https://alcf.anl.gov/science/projects/nwchemex-tackling-chemical-materials-biochemical-challenges-exascale-era}{NWChemEx} & SYCL \\ \hline
ESP & \href{https://alcf.anl.gov/science/projects/exascale-computational-catalysis}{Catalysis} & SYCL \\ \hline
ESP & \href{https://alcf.anl.gov/science/projects/dark-sky-mining}{DarkSkyML} & AI Frameworks \\ \hline
ESP & \href{https://alcf.anl.gov/science/projects/simulating-and-learning-atlas-detector-exascale}{ATLAS} & SYCL \\ \hline
ESP & \href{https://alcf.anl.gov/science/projects/data-analytics-and-machine-learning-exascale-computational-fluid-dynamics}{CFDML} & AI Frameworks, libCEED, PETSc \\ \hline
ESP & \href{https://alcf.anl.gov/science/projects/extreme-scale-situ-visualization-and-analysis-fluid-structure-interaction}{Multiphysics} & SYCL, Kokkos \\ \hline
ESP & \href{https://alcf.anl.gov/science/projects/machine-learning-lattice-quantum-chromodynamics}{LqcdML} & AI Frameworks, QUDA, Grid \\ \hline
ESP & \href{https://alcf.anl.gov/science/projects/enabling-connectomics-exascale-facilitate-discoveries-neuroscience}{Connectomics} & AI Frameworks \\ \hline
ESP & \href{https://alcf.anl.gov/science/projects/many-body-perturbation-theory-meets-machine-learning-discover-singlet-fission}{MatML} & OpenMP \\ \hline
ESP & \href{https://alcf.anl.gov/science/projects/virtual-drug-response-prediction}{VirtualDrugDesign} & AI Frameworks \\ \hline
ESP & \href{https://alcf.anl.gov/science/projects/accelerated-deep-learning-discovery-fusion-energy-science}{FusionDL} & AI Frameworks \\ \hline
\end{tabular}
\end{table}

\begin{table}[htbp!]
\label{tab:ALCCProjectProgrammingApproachChoices}
\caption {Aurora programming choice approaches for year-one (2024-2025) ALCC projects.}
\begin{tabular}{|l|p{0.25\textwidth}|>{\RaggedRight}p{0.55\textwidth}|}
\hline
\textbf{Program} & \textbf{Project} & \textbf{Approaches} \\ \hline
ALCC & \href{https://alcf.anl.gov/science/projects/building-digital-twin-model-host-pathogen-system-enhancing-biopreparedness}{TwinHostPath} & SYCL, OpenCL \\ \hline
ALCC & \href{https://alcf.anl.gov/science/projects/foundation-neuroscience-ai-model-neurox}{NeuroX} & AI Frameworks \\ \hline
ALCC & \href{https://alcf.anl.gov/science/projects/high-fidelity-cfd-enabling-advanced-nuclear-power}{CFDReac} & OCCA \\ \hline
ALCC & \href{https://alcf.anl.gov/science/projects/dns-buoyancy-driven-flows-developing-nn-informed-high-fidelity-turbulence-closures}{BuoyDrivenFlows} & OCCA \\ \hline
ALCC & \href{https://alcf.anl.gov/science/projects/exascale-computing-energy-applications}{EnergyApps} & OCCA \\ \hline
ALCC & \href{https://alcf.anl.gov/science/projects/scalable-and-resilient-modeling-federated-learning-systems-and-applications}{SR\_APPFL} & AI Frameworks \\ \hline
ALCC & \href{https://alcf.anl.gov/science/projects/autonomy-doe-simulations}{AutoSim} & AI Frameworks \\ \hline
ALCC & \href{https://alcf.anl.gov/science/projects/hadronic-contributions-muon-g-2-lattice-qcd-0}{HadronicGMinus2} & QUDA, Grid \\ \hline
ALCC & \href{https://alcf.anl.gov/science/projects/exploring-exascale-quantum-chemical-methods-transition-metal-chemistry}{TransMetalChem} & SYCL \\ \hline
ALCC & \href{https://alcf.anl.gov/science/projects/highly-scalable-ab-initio-simulations-n-doped-porous-materials-carbon-capture}{PorousMatCarbon} & OpenMP \\ \hline
ALCC & \href{https://alcf.anl.gov/science/projects/high-energy-density-physics-novel-inertial-fusion-energy-ablator-materials}{FusAblator} & Kokkos \\ \hline
ALCC & \href{https://alcf.anl.gov/science/projects/simulating-large-scale-long-lived-neutron-star-remnants-binary-neutron-star}{NeutronStarRemnants} & HIP \\ \hline
ALCC & \href{https://alcf.anl.gov/science/projects/predicting-heterogeneous-photocatalysts-using-large-scale-ab-initio-calculations}{PredictPhotocatal} & OpenMP \\ \hline
\end{tabular}
\end{table}

\begin{table}[htbp!]
\label{tab:INCITEProjectProgrammingApproachChoices}
\caption {Aurora programming choice approaches for year-one INCITE projects.}
\begin{tabular}{|l|p{0.25\textwidth}|>{\RaggedRight}p{0.55\textwidth}|}
\hline
\textbf{Program} & \textbf{Project} & \textbf{Approaches} \\ \hline
INCITE & \href{https://alcf.anl.gov/science/projects/state-art-high-resolution-3d-simulations-core-collapse-supernovae-0}{CoreCollapseModel} & OpenMP \\ \hline
INCITE & \href{https://alcf.anl.gov/science/projects/foundation-models-predictive-molecular-epidemiology-0}{FoundEpidem} & AI Frameworks \\ \hline
INCITE & \href{https://alcf.anl.gov/science/projects/heteropolymer-design-harnessing-new-and-emerging-computing-technologies-0}{HetPolyDesign} & OpenMP, SYCL \\ \hline
INCITE & \href{https://alcf.anl.gov/science/projects/advances-quark-and-lepton-flavor-physics-lattice-qcd-0}{LatticeFlavor} & SYCL, QUDA \\ \hline
INCITE & \href{https://alcf.anl.gov/science/projects/radiation-dominated-black-hole-accretion-1}{RadBlackHoleAcc} & Kokkos \\ \hline
INCITE & \href{https://alcf.anl.gov/science/projects/probing-primordial-universe-exascale-simulations}{ProbingPrimordialU} & OpenMP, SYCL \\ \hline
INCITE & \href{https://alcf.anl.gov/science/projects/multi-resolution-genome-folding-ensemble-3d-structures-across-diverse-tissues}{MRGenomeFolding} & OpenMP \\ \hline
INCITE & \href{https://alcf.anl.gov/science/projects/energy-exascale-earth-system-model-2}{E3SM\_Dec} & Kokkos \\ \hline
INCITE & \href{https://alcf.anl.gov/science/projects/exascale-gyrokinetic-study-iter-challenge-power-exhaust-and-elm-free-edge-1}{TokamakITER} & Kokkos \\ \hline
INCITE & \href{https://alcf.anl.gov/science/projects/ai-guided-exascale-simulations-quantum-materials-manufacturing-and-control-1}{QuantMatManufact} & OpenMP, AI Frameworks \\ \hline
INCITE & \href{https://alcf.anl.gov/science/projects/combining-deep-learning-physics-based-affinity-estimation-3-compbio3-0}{CompBioAffin} & SYCL \\ \hline
INCITE & \href{https://alcf.anl.gov/science/projects/auroragpt-large-scale-foundation-model-advancing-science}{AuroraGPT} & AI Frameworks \\ \hline
INCITE & \href{https://alcf.anl.gov/science/projects/qmc-hamm-high-accuracy-multiscale-models-using-quantum-monte-carlo}{qmchamm} & OpenMP, SYCL, Kokkos \\ \hline
INCITE & \href{https://alcf.anl.gov/science/projects/exascale-simulations-rayleigh-benard-convection}{RBC\_Conv\_2} & OCCA \\ \hline
INCITE & \href{https://alcf.anl.gov/science/projects/exascale-simulations-quantum-materials-1}{PSFMat\_2} & OpenMP, SYCL \\ \hline
INCITE & \href{https://alcf.anl.gov/science/projects/3d-imaging-strong-interaction-nambu-goldstone-bosons-0}{StructNGB} & Grid \\ \hline
INCITE & \href{https://alcf.anl.gov/science/projects/ab-initio-nuclear-structure-and-nuclear-reactions-4}{NucStructReact\_6} & OpenMP \\ \hline
INCITE & \href{https://alcf.anl.gov/science/projects/exascale-simulations-compact-binary-mergers}{CompactBinaryMerger} & Kokkos \\ \hline
INCITE & \href{https://alcf.anl.gov/science/projects/using-exascale-computing-new-insights-high-lift-ground-testing}{HighLiftGroundTest} & SYCL \\ \hline
INCITE & \href{https://alcf.anl.gov/science/projects/exascale-catalytic-chemistry-0}{ExaCatChem} & OpenMP, SYCL \\ \hline
INCITE & \href{https://alcf.anl.gov/science/projects/qcd-under-extreme-conditions-0}{ExtremeQCD} & \textit{TBD} \\ \hline
INCITE & \href{https://alcf.anl.gov/science/projects/mofa-generative-ai-driven-mof-discovery-carbon-capture-exascale}{MOFA} & Kokkos, AI Frameworks, HIP \\ \hline
INCITE & \href{https://alcf.anl.gov/science/projects/online-machine-learning-large-scale-turbulent-simulations-1}{OnlineTurb} & AI Frameworks, libCEED, PETSc \\ \hline
INCITE & \href{https://alcf.anl.gov/science/projects/experimental-realization-certified-randomness-quantum-supremacy}{CertRandomness} & AI Frameworks \\ \hline
INCITE & \href{https://alcf.anl.gov/science/projects/carbon-extremes-discovery-science-exascale-computers-0}{ExtremeCarbon} & Kokkos \\ \hline
INCITE & \href{https://alcf.anl.gov/science/projects/exacortex-exascale-reconstruction-human-cerebral-cortex-0}{ExaCortex} & AI Frameworks \\ \hline
INCITE & \href{https://alcf.anl.gov/science/projects/establishing-digital-twins-high-throughput-cellular-analysis-whole-blood-0}{TwinBlood} & AI Frameworks \\ \hline
INCITE & \href{https://alcf.anl.gov/science/projects/advancing-fusion-and-fission-energy-through-exascale}{AdvanceFusionFission} & SYCL, Kokkos \\ \hline
INCITE & \href{https://alcf.anl.gov/science/projects/high-accuracy-quantum-simulations-cancer-therapy-using-exascale-computing}{QSimCancerTherapy} & OpenMP, OCCA \\ \hline
INCITE & \href{https://alcf.anl.gov/science/projects/climate-informed-large-scale-and-high-resolution-inundation-modeling-framework}{InundationModel} & \textit{TBD} \\ \hline
INCITE & \href{https://alcf.anl.gov/science/projects/exascale-simulation-and-deep-learning-model-energetic-particles-burning-plasmas}{DLBurningPlasmas} & AI Frameworks, \textit{TBD} \\ \hline
INCITE & \href{https://alcf.anl.gov/science/projects/exascale-simulation-correlated-election-phonon-coupling-quantum-materials}{ElectronPhonon} & OpenMP \\ \hline
INCITE & \href{https://alcf.anl.gov/science/projects/heterogeneous-reaction-dynamics-energy-storage-and-hydrogen-production-0}{HetRxnEnergy} & Kokkos, SYCL, AI Frameworks, Kokkos-kernels \\ \hline
INCITE & \href{https://alcf.anl.gov/science/projects/using-exascale-computing-new-insights-high-lift-ground-testing}{OpenFanWing} & SYCL \\ \hline
\end{tabular}
\end{table}

\begin{table}[htbp!]
\label{tab:ECPProjectProgrammingApproachChoices}
\caption {Aurora programming choice approaches for ECP AD projects.}
\begin{tabular}{|l|p{0.25\textwidth}|>{\RaggedRight}p{0.55\textwidth}|}
\hline
\textbf{Program} & \textbf{Project} & \textbf{Approaches} \\ \hline
ECP & WDMApp & OpenMP, SYCL, Kokkos \\ \hline
ECP & ExaWind & Kokkos, AMReX \\ \hline
ECP & EXAALT & OpenMP, Kokkos \\ \hline
ECP & ExaStar & OpenMP \\ \hline
ECP & GAMESS & OpenMP, HIP \\ \hline
ECP & NWChemEx & SYCL \\ \hline
ECP & ExaSky & SYCL, AMReX \\ \hline
ECP & LatticeQCD & QUDA, Grid \\ \hline
ECP & CANDLE & AI Frameworks \\ \hline
ECP & E3SM-MMF & SYCL, Kokkos \\ \hline
ECP & EQSim & RAJA \\ \hline
ECP & MFIX-Exa & AMReX \\ \hline
ECP & QMCPACK & OpenMP \\ \hline
ECP & ExaSMR & OpenMP, OCCA \\ \hline
\end{tabular}
\end{table}

\end{appendices}

\clearpage

\bibliography{references}


\begin{thebibliography}{95}
\ifx \bisbn   \undefined \def \bisbn  #1{ISBN #1}\fi
\ifx \binits  \undefined \def \binits#1{#1}\fi
\ifx \bauthor  \undefined \def \bauthor#1{#1}\fi
\ifx \batitle  \undefined \def \batitle#1{#1}\fi
\ifx \bjtitle  \undefined \def \bjtitle#1{#1}\fi
\ifx \bvolume  \undefined \def \bvolume#1{\textbf{#1}}\fi
\ifx \byear  \undefined \def \byear#1{#1}\fi
\ifx \bissue  \undefined \def \bissue#1{#1}\fi
\ifx \bfpage  \undefined \def \bfpage#1{#1}\fi
\ifx \blpage  \undefined \def \blpage #1{#1}\fi
\ifx \burl  \undefined \def \burl#1{\textsf{#1}}\fi
\ifx \doiurl  \undefined \def \doiurl#1{\url{https://doi.org/#1}}\fi
\ifx \betal  \undefined \def \betal{\textit{et al.}}\fi
\ifx \binstitute  \undefined \def \binstitute#1{#1}\fi
\ifx \binstitutionaled  \undefined \def \binstitutionaled#1{#1}\fi
\ifx \bctitle  \undefined \def \bctitle#1{#1}\fi
\ifx \beditor  \undefined \def \beditor#1{#1}\fi
\ifx \bpublisher  \undefined \def \bpublisher#1{#1}\fi
\ifx \bbtitle  \undefined \def \bbtitle#1{#1}\fi
\ifx \bedition  \undefined \def \bedition#1{#1}\fi
\ifx \bseriesno  \undefined \def \bseriesno#1{#1}\fi
\ifx \blocation  \undefined \def \blocation#1{#1}\fi
\ifx \bsertitle  \undefined \def \bsertitle#1{#1}\fi
\ifx \bsnm \undefined \def \bsnm#1{#1}\fi
\ifx \bsuffix \undefined \def \bsuffix#1{#1}\fi
\ifx \bparticle \undefined \def \bparticle#1{#1}\fi
\ifx \barticle \undefined \def \barticle#1{#1}\fi
\bibcommenthead
\ifx \bconfdate \undefined \def \bconfdate #1{#1}\fi
\ifx \botherref \undefined \def \botherref #1{#1}\fi
\ifx \url \undefined \def \url#1{\textsf{#1}}\fi
\ifx \bchapter \undefined \def \bchapter#1{#1}\fi
\ifx \bbook \undefined \def \bbook#1{#1}\fi
\ifx \bcomment \undefined \def \bcomment#1{#1}\fi
\ifx \oauthor \undefined \def \oauthor#1{#1}\fi
\ifx \citeauthoryear \undefined \def \citeauthoryear#1{#1}\fi
\ifx \endbibitem  \undefined \def \endbibitem {}\fi
\ifx \bconflocation  \undefined \def \bconflocation#1{#1}\fi
\ifx \arxivurl  \undefined \def \arxivurl#1{\textsf{#1}}\fi
\csname PreBibitemsHook\endcsname

\bibitem[\protect\citeauthoryear{}{}]{TOP500}
\begin{botherref}
{TOP500} List.
\url{https://www.top500.org/lists/top500/}
Accessed 26 March 2025
\end{botherref}
\endbibitem

\bibitem[\protect\citeauthoryear{Gomes et~al.}{2022}]{9731673}
\begin{bchapter}
\bauthor{\bsnm{Gomes}, \binits{W.}},
\bauthor{\bsnm{Koker}, \binits{A.}},
\bauthor{\bsnm{Stover}, \binits{P.}},
\bauthor{\bsnm{Ingerly}, \binits{D.}},
\bauthor{\bsnm{Siers}, \binits{S.}},
\bauthor{\bsnm{Venkataraman}, \binits{S.}},
\bauthor{\bsnm{Pelto}, \binits{C.}},
\bauthor{\bsnm{Shah}, \binits{T.}},
\bauthor{\bsnm{Rao}, \binits{A.}},
\bauthor{\bsnm{O'Mahony}, \binits{F.}},
\bauthor{\bsnm{Karl}, \binits{E.}},
\bauthor{\bsnm{Cheney}, \binits{L.}},
\bauthor{\bsnm{Rajwani}, \binits{I.}},
\bauthor{\bsnm{Jain}, \binits{H.}},
\bauthor{\bsnm{Cortez}, \binits{R.}},
\bauthor{\bsnm{Chandrasekhar}, \binits{A.}},
\bauthor{\bsnm{Kanthi}, \binits{B.}},
\bauthor{\bsnm{Koduri}, \binits{R.}}:
\bctitle{{Ponte Vecchio: A Multi-Tile 3D Stacked Processor for Exascale Computing}}.
In: \bbtitle{2022 IEEE International Solid-State Circuits Conference (ISSCC)},
vol. \bseriesno{65},
pp. \bfpage{42}--\blpage{44}
(\byear{2022}).
\doiurl{10.1109/ISSCC42614.2022.9731673}
\end{bchapter}
\endbibitem

\bibitem[\protect\citeauthoryear{Nassif et~al.}{2022}]{9731107}
\begin{bchapter}
\bauthor{\bsnm{Nassif}, \binits{N.}},
\bauthor{\bsnm{Munch}, \binits{A.O.}},
\bauthor{\bsnm{Molnar}, \binits{C.L.}},
\bauthor{\bsnm{Pasdast}, \binits{G.}},
\bauthor{\bsnm{Lyer}, \binits{S.V.}},
\bauthor{\bsnm{Yang}, \binits{Z.}},
\bauthor{\bsnm{Mendoza}, \binits{O.}},
\bauthor{\bsnm{Huddart}, \binits{M.}},
\bauthor{\bsnm{Venkataraman}, \binits{S.}},
\bauthor{\bsnm{Kandula}, \binits{S.}},
\bauthor{\bsnm{Marom}, \binits{R.}},
\bauthor{\bsnm{Kern}, \binits{A.M.}},
\bauthor{\bsnm{Bowhill}, \binits{B.}},
\bauthor{\bsnm{Mulvihill}, \binits{D.R.}},
\bauthor{\bsnm{Nimmagadda}, \binits{S.}},
\bauthor{\bsnm{Kalidindi}, \binits{V.}},
\bauthor{\bsnm{Krause}, \binits{J.}},
\bauthor{\bsnm{Haq}, \binits{M.M.}},
\bauthor{\bsnm{Sharma}, \binits{R.}},
\bauthor{\bsnm{Duda}, \binits{K.}}:
\bctitle{{Sapphire Rapids: The Next-Generation Intel Xeon Scalable Processor}}.
In: \bbtitle{2022 IEEE International Solid-State Circuits Conference (ISSCC)},
vol. \bseriesno{65},
pp. \bfpage{44}--\blpage{46}
(\byear{2022}).
\doiurl{10.1109/ISSCC42614.2022.9731107}
\end{bchapter}
\endbibitem

\bibitem[\protect\citeauthoryear{}{}]{largest_ss}
\begin{botherref}
Largest HPE Slingshot deployment.
\url{https://www.hpe.com/us/en/newsroom/press-release/2024/05/hewlett-packard-enterprise-delivers-second-exascale-supercomputer-aurora-to-us-department-of-energys-argonne-national-laboratory.html}
Accessed 26 March 2025
\end{botherref}
\endbibitem

\bibitem[\protect\citeauthoryear{Zhen et~al.}{2020}]{hennecke2020daos}
\begin{botherref}
\oauthor{\bsnm{Zhen}, \binits{L.}},
\oauthor{\bsnm{Lombardi}, \binits{J.}},
\oauthor{\bsnm{Chaarawi}, \binits{M.}},
\oauthor{\bsnm{Hennecke}, \binits{M.}}:
{Daos: A scale-out high performance storage stack for storage class memory}.
Supercomputing frontiers
\textbf{40}
(2020)
\end{botherref}
\endbibitem

\bibitem[\protect\citeauthoryear{Kim et~al.}{2018}]{Kim_2018}
\begin{barticle}
\bauthor{\bsnm{Kim}, \binits{J.}},
\bauthor{\bsnm{Baczewski}, \binits{A.D.}},
\bauthor{\bsnm{Beaudet}, \binits{T.D.}},
\bauthor{\bsnm{Benali}, \binits{A.}},
\bauthor{\bsnm{Bennett}, \binits{M.C.}},
\bauthor{\bsnm{Berrill}, \binits{M.A.}},
\bauthor{\bsnm{Blunt}, \binits{N.S.}},
\bauthor{\bsnm{Borda}, \binits{E.J.L.}},
\bauthor{\bsnm{Casula}, \binits{M.}},
\bauthor{\bsnm{Ceperley}, \binits{D.M.}},
\bauthor{\bsnm{Chiesa}, \binits{S.}},
\bauthor{\bsnm{Clark}, \binits{B.K.}},
\bauthor{\bsnm{Clay}, \binits{R.C.}},
\bauthor{\bsnm{Delaney}, \binits{K.T.}},
\bauthor{\bsnm{Dewing}, \binits{M.}},
\bauthor{\bsnm{Esler}, \binits{K.P.}},
\bauthor{\bsnm{Hao}, \binits{H.}},
\bauthor{\bsnm{Heinonen}, \binits{O.}},
\bauthor{\bsnm{Kent}, \binits{P.R.C.}},
\bauthor{\bsnm{Krogel}, \binits{J.T.}},
\bauthor{\bsnm{Kylänpää}, \binits{I.}},
\bauthor{\bsnm{Li}, \binits{Y.W.}},
\bauthor{\bsnm{Lopez}, \binits{M.G.}},
\bauthor{\bsnm{Luo}, \binits{Y.}},
\bauthor{\bsnm{Malone}, \binits{F.D.}},
\bauthor{\bsnm{Martin}, \binits{R.M.}},
\bauthor{\bsnm{Mathuriya}, \binits{A.}},
\bauthor{\bsnm{McMinis}, \binits{J.}},
\bauthor{\bsnm{Melton}, \binits{C.A.}},
\bauthor{\bsnm{Mitas}, \binits{L.}},
\bauthor{\bsnm{Morales}, \binits{M.A.}},
\bauthor{\bsnm{Neuscamman}, \binits{E.}},
\bauthor{\bsnm{Parker}, \binits{W.D.}},
\bauthor{\bsnm{Pineda~Flores}, \binits{S.D.}},
\bauthor{\bsnm{Romero}, \binits{N.A.}},
\bauthor{\bsnm{Rubenstein}, \binits{B.M.}},
\bauthor{\bsnm{Shea}, \binits{J.A.R.}},
\bauthor{\bsnm{Shin}, \binits{H.}},
\bauthor{\bsnm{Shulenburger}, \binits{L.}},
\bauthor{\bsnm{Tillack}, \binits{A.F.}},
\bauthor{\bsnm{Townsend}, \binits{J.P.}},
\bauthor{\bsnm{Tubman}, \binits{N.M.}},
\bauthor{\bsnm{Van Der~Goetz}, \binits{B.}},
\bauthor{\bsnm{Vincent}, \binits{J.E.}},
\bauthor{\bsnm{Yang}, \binits{D.C.}},
\bauthor{\bsnm{Yang}, \binits{Y.}},
\bauthor{\bsnm{Zhang}, \binits{S.}},
\bauthor{\bsnm{Zhao}, \binits{L.}}:
\batitle{Qmcpack: an open source ab initio quantum monte carlo package for the electronic structure of atoms, molecules and solids}.
\bjtitle{Journal of Physics: Condensed Matter}
\bvolume{30}(\bissue{19}),
\bfpage{195901}
(\byear{2018})
\doiurl{10.1088/1361-648X/aab9c3}
\end{barticle}
\endbibitem

\bibitem[\protect\citeauthoryear{Berzins et~al.}{2010}]{Uintah}
\begin{bchapter}
\bauthor{\bsnm{Berzins}, \binits{M.}},
\bauthor{\bsnm{Luitjens}, \binits{J.}},
\bauthor{\bsnm{Meng}, \binits{Q.}},
\bauthor{\bsnm{Harman}, \binits{T.}},
\bauthor{\bsnm{Wight}, \binits{C.A.}},
\bauthor{\bsnm{Peterson}, \binits{J.R.}}:
\bctitle{Uintah: a scalable framework for hazard analysis}.
In: \bbtitle{Proceedings of the 2010 TeraGrid Conference}.
\bsertitle{TG '10}.
\bpublisher{Association for Computing Machinery},
\blocation{New York, NY, USA}
(\byear{2010}).
\doiurl{10.1145/1838574.1838577}
\end{bchapter}
\endbibitem

\bibitem[\protect\citeauthoryear{Ku et~al.}{2018a}]{XGC1}
\begin{botherref}
\oauthor{\bsnm{Ku}, \binits{S.-H.}},
\oauthor{\bsnm{Hager}, \binits{R.}},
\oauthor{\bsnm{Scheinberg}, \binits{A.}},
\oauthor{\bsnm{Dominski}, \binits{J.}},
\oauthor{\bsnm{Sharma}, \binits{A.}},
\oauthor{\bsnm{Churchill}, \binits{M.}},
\oauthor{\bsnm{Choi}, \binits{J.}},
\oauthor{\bsnm{Sturdevant}, \binits{B.}},
\oauthor{\bsnm{Mollén}, \binits{A.}},
\oauthor{\bsnm{Wilkie}, \binits{G.}},
\oauthor{\bsnm{Chang}, \binits{C.-S.}},
\oauthor{\bsnm{Yoon}, \binits{E.}},
\oauthor{\bsnm{Adams}, \binits{M.}},
\oauthor{\bsnm{Seo}, \binits{J.}},
\oauthor{\bsnm{Koh}, \binits{S.}},
\oauthor{\bsnm{D'Azevedo}, \binits{E.}},
\oauthor{\bsnm{Abbott}, \binits{S.}},
\oauthor{\bsnm{Worley}, \binits{P.H.}},
\oauthor{\bsnm{Ethier}, \binits{S.}},
\oauthor{\bsnm{Park}, \binits{G.}},
\oauthor{\bsnm{Lang}, \binits{J.}},
\oauthor{\bsnm{MacKie-Mason}, \binits{B.}},
\oauthor{\bsnm{Germaschewski}, \binits{K.}},
\oauthor{\bsnm{Suchyta}, \binits{E.}},
\oauthor{\bsnm{Carey}, \binits{V.}},
\oauthor{\bsnm{Cole}, \binits{M.}},
\oauthor{\bsnm{Trivedi}, \binits{P.}},
\oauthor{\bsnm{Chowdhury}, \binits{J.}}:
XGC.
[Computer Software] \url{https://doi.org/10.11578/dc.20180627.11}
(2018).
\doiurl{10.11578/dc.20180627.11}
\end{botherref}
\endbibitem

\bibitem[\protect\citeauthoryear{Ku et~al.}{2018b}]{XGC2}
\begin{barticle}
\bauthor{\bsnm{Ku}, \binits{S.}},
\bauthor{\bsnm{Chang}, \binits{C.S.}},
\bauthor{\bsnm{Hager}, \binits{R.}},
\bauthor{\bsnm{Churchill}, \binits{R.M.}},
\bauthor{\bsnm{Tynan}, \binits{G.R.}},
\bauthor{\bsnm{Cziegler}, \binits{I.}},
\bauthor{\bsnm{Greenwald}, \binits{M.}},
\bauthor{\bsnm{Hughes}, \binits{J.}},
\bauthor{\bsnm{Parker}, \binits{S.E.}},
\bauthor{\bsnm{Adams}, \binits{M.F.}},
\bauthor{\bsnm{D'Azevedo}, \binits{E.}},
\bauthor{\bsnm{Worley}, \binits{P.}}:
\batitle{A fast low-to-high confinement mode bifurcation dynamics in the boundary-plasma gyrokinetic code xgc1}.
\bjtitle{Physics of Plasmas}
\bvolume{25}(\bissue{5}),
\bfpage{056107}
(\byear{2018})
\doiurl{10.1063/1.5020792}
{\href{https://arxiv.org/abs/https://pubs.aip.org/aip/pop/article-pdf/doi/10.1063/1.5020792/19756501/056107\_1\_online.pdf}{{https://pubs.aip.org/aip/pop/article-pdf/doi/10.1063/1.5020792/19756501/056107\_1\_online.pdf}}}
\end{barticle}
\endbibitem

\bibitem[\protect\citeauthoryear{Hager et~al.}{2022}]{XGC3}
\begin{barticle}
\bauthor{\bsnm{Hager}, \binits{R.}},
\bauthor{\bsnm{Ku}, \binits{S.}},
\bauthor{\bsnm{Sharma}, \binits{A.Y.}},
\bauthor{\bsnm{Chang}, \binits{C.S.}},
\bauthor{\bsnm{Churchill}, \binits{R.M.}},
\bauthor{\bsnm{Scheinberg}, \binits{A.}}:
\batitle{Electromagnetic total-f algorithm for gyrokinetic particle-in-cell simulations of boundary plasma in xgc}.
\bjtitle{Physics of Plasmas}
\bvolume{29}(\bissue{11}),
\bfpage{112308}
(\byear{2022})
\doiurl{10.1063/5.0097855}
{\href{https://arxiv.org/abs/https://pubs.aip.org/aip/pop/article-pdf/doi/10.1063/5.0097855/16627866/112308\_1\_online.pdf}{{https://pubs.aip.org/aip/pop/article-pdf/doi/10.1063/5.0097855/16627866/112308\_1\_online.pdf}}}
\end{barticle}
\endbibitem

\bibitem[\protect\citeauthoryear{}{}]{MILC}
\begin{botherref}
MILC code repository.
\url{https://github.com/milc-qcd/milc_qcd}
Accessed 26 March 2025
\end{botherref}
\endbibitem

\bibitem[\protect\citeauthoryear{}{}]{Chroma}
\begin{botherref}
Chroma code repository.
\url{https://jeffersonlab.github.io/chroma}
Accessed 26 March 2025
\end{botherref}
\endbibitem

\bibitem[\protect\citeauthoryear{}{}]{Grid}
\begin{botherref}
Grid code repository.
\url{https://github.com/paboyle/Grid}
Accessed 26 March 2025
\end{botherref}
\endbibitem

\bibitem[\protect\citeauthoryear{}{}]{QUDA}
\begin{botherref}
QUDA code repository.
\url{https://lattice.github.io/quda}
Accessed 26 March 2025
\end{botherref}
\endbibitem

\bibitem[\protect\citeauthoryear{Habib et~al.}{2016}]{HABIB201649}
\begin{barticle}
\bauthor{\bsnm{Habib}, \binits{S.}},
\bauthor{\bsnm{Pope}, \binits{A.}},
\bauthor{\bsnm{Finkel}, \binits{H.}},
\bauthor{\bsnm{Frontiere}, \binits{N.}},
\bauthor{\bsnm{Heitmann}, \binits{K.}},
\bauthor{\bsnm{Daniel}, \binits{D.}},
\bauthor{\bsnm{Fasel}, \binits{P.}},
\bauthor{\bsnm{Morozov}, \binits{V.}},
\bauthor{\bsnm{Zagaris}, \binits{G.}},
\bauthor{\bsnm{Peterka}, \binits{T.}},
\bauthor{\bsnm{Vishwanath}, \binits{V.}},
\bauthor{\bsnm{Lukić}, \binits{Z.}},
\bauthor{\bsnm{Sehrish}, \binits{S.}},
\bauthor{\bsnm{Liao}, \binits{W.-k.}}:
\batitle{Hacc: Simulating sky surveys on state-of-the-art supercomputing architectures}.
\bjtitle{New Astronomy}
\bvolume{42},
\bfpage{49}--\blpage{65}
(\byear{2016})
\doiurl{10.1016/j.newast.2015.06.003}
\end{barticle}
\endbibitem

\bibitem[\protect\citeauthoryear{Whiting and Jansen}{1999}]{PHASTA1}
\begin{botherref}
\oauthor{\bsnm{Whiting}, \binits{C.H.}},
\oauthor{\bsnm{Jansen}, \binits{K.E.}}:
Stabilized finite element methods for fluid dynamics using a hierarchical basis.
PhD thesis,
Rensselaer Polytechnic Institute
(1999)
\end{botherref}
\endbibitem

\bibitem[\protect\citeauthoryear{Jansen et~al.}{2000}]{PHASTA2}
\begin{barticle}
\bauthor{\bsnm{Jansen}, \binits{K.E.}},
\bauthor{\bsnm{Whiting}, \binits{C.H.}},
\bauthor{\bsnm{Hulbert}, \binits{G.M.}}:
\batitle{{Generalized-$\alpha$ method for integrating the filtered Navier-Stokes equations with a stabilized finite element method}}.
\bjtitle{Computer Methods in Applied Mechanics and Engineering}
(\byear{2000})
\doiurl{10.1016/S0045-7825(00)00203-6}
\end{barticle}
\endbibitem

\bibitem[\protect\citeauthoryear{Phillips et~al.}{2020}]{NAMD}
\begin{barticle}
\bauthor{\bsnm{Phillips}, \binits{J.C.}},
\bauthor{\bsnm{Hardy}, \binits{D.J.}},
\bauthor{\bsnm{Maia}, \binits{J.D.C.}},
\bauthor{\bsnm{Stone}, \binits{J.E.}},
\bauthor{\bsnm{Ribeiro}, \binits{J.V.}},
\bauthor{\bsnm{Bernardi}, \binits{R.C.}},
\bauthor{\bsnm{Buch}, \binits{R.}},
\bauthor{\bsnm{Fiorin}, \binits{G.}},
\bauthor{\bsnm{Hénin}, \binits{J.}},
\bauthor{\bsnm{Jiang}, \binits{W.}},
\bauthor{\bsnm{McGreevy}, \binits{R.}},
\bauthor{\bsnm{Melo}, \binits{M.C.R.}},
\bauthor{\bsnm{Radak}, \binits{B.K.}},
\bauthor{\bsnm{Skeel}, \binits{R.D.}},
\bauthor{\bsnm{Singharoy}, \binits{A.}},
\bauthor{\bsnm{Wang}, \binits{Y.}},
\bauthor{\bsnm{Roux}, \binits{B.}},
\bauthor{\bsnm{Aksimentiev}, \binits{A.}},
\bauthor{\bsnm{Luthey-Schulten}, \binits{Z.}},
\bauthor{\bsnm{Kalé}, \binits{L.V.}},
\bauthor{\bsnm{Schulten}, \binits{K.}},
\bauthor{\bsnm{Chipot}, \binits{C.}},
\bauthor{\bsnm{Tajkhorshid}, \binits{E.}}:
\batitle{Scalable molecular dynamics on cpu and gpu architectures with namd}.
\bjtitle{The Journal of Chemical Physics}
\bvolume{153}(\bissue{4}),
\bfpage{044130}
(\byear{2020})
\doiurl{10.1063/5.0014475}
{\href{https://arxiv.org/abs/https://pubs.aip.org/aip/jcp/article-pdf/doi/10.1063/5.0014475/16709546/044130\_1\_online.pdf}{{https://pubs.aip.org/aip/jcp/article-pdf/doi/10.1063/5.0014475/16709546/044130\_1\_online.pdf}}}
\end{barticle}
\endbibitem

\bibitem[\protect\citeauthoryear{Kowalski et~al.}{2021}]{nwchemex}
\begin{barticle}
\bauthor{\bsnm{Kowalski}, \binits{K.}},
\bauthor{\bsnm{Bair}, \binits{R.}},
\bauthor{\bsnm{Bauman}, \binits{N.P.}},
\bauthor{\bsnm{Boschen}, \binits{J.S.}},
\bauthor{\bsnm{Bylaska}, \binits{E.J.}},
\bauthor{\bsnm{Daily}, \binits{J.}},
\bauthor{\bsnm{Jong}, \binits{W.A.}},
\bauthor{\bsnm{Dunning~Jr}, \binits{T.}},
\bauthor{\bsnm{Govind}, \binits{N.}},
\bauthor{\bsnm{Harrison}, \binits{R.J.}}, \betal:
\batitle{From nwchem to nwchemex: Evolving with the computational chemistry landscape}.
\bjtitle{Chemical reviews}
\bvolume{121}(\bissue{8}),
\bfpage{4962}--\blpage{4998}
(\byear{2021})
\end{barticle}
\endbibitem

\bibitem[\protect\citeauthoryear{Bylaska et~al.}{2024}]{BYLASKA2024518}
\begin{bchapter}
\bauthor{\bsnm{Bylaska}, \binits{E.J.}},
\bauthor{\bsnm{Song}, \binits{D.}},
\bauthor{\bsnm{Ilton}, \binits{E.S.}},
\bauthor{\bsnm{Bagusetty}, \binits{A.}},
\bauthor{\bsnm{Bross}, \binits{D.}},
\bauthor{\bsnm{Vazquez-Mayagoitia}, \binits{A.}},
\bauthor{\bsnm{Hernandez}, \binits{R.}},
\bauthor{\bsnm{Gawande}, \binits{N.}}:
\bctitle{Nwchem and nwchemex plane-wave methods}.
In: \beditor{\bsnm{Yáñez}, \binits{M.}},
\beditor{\bsnm{Boyd}, \binits{R.J.}} (eds.)
\bbtitle{Comprehensive Computational Chemistry (First Edition)},
\bedition{1}st edn.,
pp. \bfpage{518}--\blpage{543}.
\bpublisher{Elsevier},
\blocation{Oxford}
(\byear{2024}).
\doiurl{10.1016/B978-0-12-821978-2.00094-5} .
\burl{https://www.sciencedirect.com/science/article/pii/B9780128219782000945}
\end{bchapter}
\endbibitem

\bibitem[\protect\citeauthoryear{{Atif, Mohammad} et~al.}{2024}]{FastCaloSim}
\begin{barticle}
\bauthor{\bsnm{{Atif, Mohammad}}},
\bauthor{\bsnm{{Dong, Zhihua}}},
\bauthor{\bsnm{{Leggett, Charles}}},
\bauthor{\bsnm{{Lin, Meifeng}}},
\bauthor{\bsnm{{Tsulaia, Vakhtang}}}:
\batitle{Porting atlas fast calorimeter simulation to gpus with performance portable programming models}.
\bjtitle{EPJ Web of Conf.}
\bvolume{295},
\bfpage{11018}
(\byear{2024})
\doiurl{10.1051/epjconf/202429511018}
\end{barticle}
\endbibitem

\bibitem[\protect\citeauthoryear{Nichols et~al.}{2024}]{MadGraph}
\begin{bchapter}
\bauthor{\bsnm{Nichols}, \binits{N.S.}},
\bauthor{\bsnm{Childers}, \binits{J.T.}},
\bauthor{\bsnm{Burch}, \binits{T.J.}},
\bauthor{\bsnm{Field}, \binits{L.}}:
\bctitle{Improving performance portability of the procedurally generated high energy physics event generator madgraph using sycl}.
In: \bbtitle{Proceedings of the 12th International Workshop on OpenCL and SYCL}.
\bsertitle{IWOCL '24}.
\bpublisher{Association for Computing Machinery},
\blocation{New York, NY, USA}
(\byear{2024}).
\doiurl{10.1145/3648115.3648116}
\end{bchapter}
\endbibitem

\bibitem[\protect\citeauthoryear{Balin et~al.}{2023}]{cfdml}
\begin{botherref}
\oauthor{\bsnm{Balin}, \binits{R.}},
\oauthor{\bsnm{Simini}, \binits{F.}},
\oauthor{\bsnm{Simpson}, \binits{C.}},
\oauthor{\bsnm{Shao}, \binits{A.}},
\oauthor{\bsnm{Rigazzi}, \binits{A.}},
\oauthor{\bsnm{Ellis}, \binits{M.}},
\oauthor{\bsnm{Becker}, \binits{S.}},
\oauthor{\bsnm{Doostan}, \binits{A.}},
\oauthor{\bsnm{Evans}, \binits{J.A.}},
\oauthor{\bsnm{Jansen}, \binits{K.E.}}:
In situ framework for coupling simulation and machine learning with application to cfd.
arXiv preprint arXiv:2306.12900
(2023)
\end{botherref}
\endbibitem

\bibitem[\protect\citeauthoryear{Randles et~al.}{2013}]{HARVEY}
\begin{bchapter}
\bauthor{\bsnm{Randles}, \binits{A.P.}},
\bauthor{\bsnm{Kale}, \binits{V.}},
\bauthor{\bsnm{Hammond}, \binits{J.}},
\bauthor{\bsnm{Gropp}, \binits{W.}},
\bauthor{\bsnm{Kaxiras}, \binits{E.}}:
\bctitle{Performance analysis of the lattice boltzmann model beyond navier-stokes}.
In: \bbtitle{Proceedings of the 2013 IEEE 27th International Symposium on Parallel and Distributed Processing}.
\bsertitle{IPDPS '13},
pp. \bfpage{1063}--\blpage{1074}.
\bpublisher{IEEE Computer Society},
\blocation{USA}
(\byear{2013}).
\doiurl{10.1109/IPDPS.2013.109}
\end{bchapter}
\endbibitem

\bibitem[\protect\citeauthoryear{Abbott et~al.}{2024}]{LQCDML}
\begin{barticle}
\bauthor{\bsnm{Abbott}, \binits{R.}},
\bauthor{\bsnm{Botev}, \binits{A.}},
\bauthor{\bsnm{Boyda}, \binits{D.}},
\bauthor{\bsnm{Hackett}, \binits{D.C.}},
\bauthor{\bsnm{Kanwar}, \binits{G.}},
\bauthor{\bsnm{Racani{\`e}re}, \binits{S.}},
\bauthor{\bsnm{Rezende}, \binits{D.J.}},
\bauthor{\bsnm{Romero-L{\'o}pez}, \binits{F.}},
\bauthor{\bsnm{Shanahan}, \binits{P.E.}},
\bauthor{\bsnm{Urban}, \binits{J.M.}}:
\batitle{Applications of flow models to the generation of correlated lattice {QCD} ensembles}.
\bjtitle{Physical Review D}
\bvolume{109}(\bissue{9}),
\bfpage{094514}
(\byear{2024})
\end{barticle}
\endbibitem

\bibitem[\protect\citeauthoryear{Shapson-Coe et~al.}{2024}]{connectomics}
\begin{barticle}
\bauthor{\bsnm{Shapson-Coe}, \binits{A.}},
\bauthor{\bsnm{Januszewski}, \binits{M.}},
\bauthor{\bsnm{Berger}, \binits{D.R.}},
\bauthor{\bsnm{Pope}, \binits{A.}},
\bauthor{\bsnm{Wu}, \binits{Y.}},
\bauthor{\bsnm{Blakely}, \binits{T.}},
\bauthor{\bsnm{Schalek}, \binits{R.L.}},
\bauthor{\bsnm{Li}, \binits{P.H.}},
\bauthor{\bsnm{Wang}, \binits{S.}},
\bauthor{\bsnm{Maitin-Shepard}, \binits{J.}},
\bauthor{\bsnm{Karlupia}, \binits{N.}},
\bauthor{\bsnm{Dorkenwald}, \binits{S.}},
\bauthor{\bsnm{Sjostedt}, \binits{E.}},
\bauthor{\bsnm{Leavitt}, \binits{L.}},
\bauthor{\bsnm{Lee}, \binits{D.}},
\bauthor{\bsnm{Troidl}, \binits{J.}},
\bauthor{\bsnm{Collman}, \binits{F.}},
\bauthor{\bsnm{Bailey}, \binits{L.}},
\bauthor{\bsnm{Fitzmaurice}, \binits{A.}},
\bauthor{\bsnm{Kar}, \binits{R.}},
\bauthor{\bsnm{Field}, \binits{B.}},
\bauthor{\bsnm{Wu}, \binits{H.}},
\bauthor{\bsnm{Wagner-Carena}, \binits{J.}},
\bauthor{\bsnm{Aley}, \binits{D.}},
\bauthor{\bsnm{Lau}, \binits{J.}},
\bauthor{\bsnm{Lin}, \binits{Z.}},
\bauthor{\bsnm{Wei}, \binits{D.}},
\bauthor{\bsnm{Pfister}, \binits{H.}},
\bauthor{\bsnm{Peleg}, \binits{A.}},
\bauthor{\bsnm{Jain}, \binits{V.}},
\bauthor{\bsnm{Lichtman}, \binits{J.W.}}:
\batitle{A petavoxel fragment of human cerebral cortex reconstructed at nanoscale resolution}.
\bjtitle{Science}
\bvolume{384}(\bissue{6696}),
\bfpage{4858}
(\byear{2024})
\doiurl{10.1126/science.adk4858}
{\href{https://arxiv.org/abs/https://www.science.org/doi/pdf/10.1126/science.adk4858}{{https://www.science.org/doi/pdf/10.1126/science.adk4858}}}
\end{barticle}
\endbibitem

\bibitem[\protect\citeauthoryear{Deslippe et~al.}{2012}]{BerkeleyGW}
\begin{barticle}
\bauthor{\bsnm{Deslippe}, \binits{J.}},
\bauthor{\bsnm{Samsonidze}, \binits{G.}},
\bauthor{\bsnm{Strubbe}, \binits{D.A.}},
\bauthor{\bsnm{Jain}, \binits{M.}},
\bauthor{\bsnm{Cohen}, \binits{M.L.}},
\bauthor{\bsnm{Louie}, \binits{S.G.}}:
\batitle{Berkeleygw: A massively parallel computer package for the calculation of the quasiparticle and optical properties of materials and nanostructures}.
\bjtitle{Computer Physics Communications}
\bvolume{183}(\bissue{6}),
\bfpage{1269}--\blpage{1289}
(\byear{2012})
\end{barticle}
\endbibitem

\bibitem[\protect\citeauthoryear{Vasan et~al.}{2023}]{SST}
\begin{bchapter}
\bauthor{\bsnm{Vasan}, \binits{A.}},
\bauthor{\bsnm{Brettin}, \binits{T.}},
\bauthor{\bsnm{Stevens}, \binits{R.}},
\bauthor{\bsnm{Ramanathan}, \binits{A.}},
\bauthor{\bsnm{Vishwanath}, \binits{V.}}:
\bctitle{Scalable lead prediction with transformers using hpc resources}.
In: \bbtitle{Proceedings of the SC '23 Workshops of the International Conference on High Performance Computing, Network, Storage, and Analysis}.
\bsertitle{SC-W '23},
p. \bfpage{123}.
\bpublisher{Association for Computing Machinery},
\blocation{New York, NY, USA}
(\byear{2023}).
\doiurl{10.1145/3624062.3624081}
\end{bchapter}
\endbibitem

\bibitem[\protect\citeauthoryear{Ibeid et~al.}{2025}]{ibeid2025hbm}
\begin{botherref}
\oauthor{\bsnm{Ibeid}, \binits{H.}},
\oauthor{\bsnm{Narayana}, \binits{V.}},
\oauthor{\bsnm{Kim}, \binits{J.}},
\oauthor{\bsnm{Nguyen}, \binits{A.}},
\oauthor{\bsnm{Morozov}, \binits{V.}},
\oauthor{\bsnm{Luo}, \binits{Y.}}:
Performance {A}nalysis of {HPC} applications on the {A}urora {S}upercomputer: {E}xploring the {I}mpact of {HBM}-{E}nabled {I}ntel {X}eon {M}ax {CPU}s
(2025).
\url{https://arxiv.org/abs/2504.03632}
\end{botherref}
\endbibitem

\bibitem[\protect\citeauthoryear{}{}]{amx}
\begin{botherref}
Optimizing Machine Learning (ML) Models with Intel® Advanced Matrix Extensions (Intel® AMX).
\url{https://www.intel.com/content/dam/www/central-libraries/us/en/documents/2022-12/optimizing-ml-models-with-amx-brief.pdf}
Accessed 26 March 2025
\end{botherref}
\endbibitem

\bibitem[\protect\citeauthoryear{Blythe}{2021}]{hotchips2021blythe}
\begin{bchapter}
\bauthor{\bsnm{Blythe}, \binits{D.}}:
\bctitle{{ XeHPC Ponte Vecchio }}.
In: \bbtitle{2021 IEEE Hot Chips 33 Symposium (HCS)},
pp. \bfpage{1}--\blpage{34}.
\bpublisher{IEEE Computer Society},
\blocation{Los Alamitos, CA, USA}
(\byear{2021}).
\doiurl{10.1109/HCS52781.2021.9567038} .
\burl{https://doi.ieeecomputersociety.org/10.1109/HCS52781.2021.9567038}
\end{bchapter}
\endbibitem

\bibitem[\protect\citeauthoryear{Jiang}{2022}]{hotchips2022jiang}
\begin{bchapter}
\bauthor{\bsnm{Jiang}, \binits{H.}}:
\bctitle{{ Intel's Ponte Vecchio GPU : Architecture, Systems \& Software }}.
In: \bbtitle{2022 IEEE Hot Chips 34 Symposium (HCS)},
pp. \bfpage{1}--\blpage{29}.
\bpublisher{IEEE Computer Society},
\blocation{Los Alamitos, CA, USA}
(\byear{2022}).
\doiurl{10.1109/HCS55958.2022.9895631} .
\burl{https://doi.ieeecomputersociety.org/10.1109/HCS55958.2022.9895631}
\end{bchapter}
\endbibitem

\bibitem[\protect\citeauthoryear{}{}]{intelgen11gfx}
\begin{botherref}
{The Architecture of Intel Processor Graphics Gen11}.
\url{https://cdrdv2-public.intel.com/686065/the-architecture-of-intel-processor-graphics-gen11-r1new.pdf}
Accessed 26 March 2025
\end{botherref}
\endbibitem

\bibitem[\protect\citeauthoryear{}{}]{intelxehpg}
\begin{botherref}
{Introduction to the Xe-HPG Architecture}.
\url{https://cdrdv2-public.intel.com/758302/introduction-to-the-xe-hpg-architecture-white-paper.pdf}
Accessed 26 March 2025
\end{botherref}
\endbibitem

\bibitem[\protect\citeauthoryear{}{}]{dma-buf}
\begin{botherref}
{Linux DMA-Buf API}.
\url{https://docs.kernel.org/driver-api/dma-buf.html}
Accessed 26 March 2025
\end{botherref}
\endbibitem

\bibitem[\protect\citeauthoryear{}{}]{pcie-p2p}
\begin{botherref}
{Linux PCIe Peer to Peer DMA}.
\url{https://docs.kernel.org/driver-api/pci/p2pdma.html}
Accessed 26 March 2025
\end{botherref}
\endbibitem

\bibitem[\protect\citeauthoryear{}{}]{shasta}
\begin{botherref}
{HPE Shasta Hardware Architecture}.
\url{https://support.hpe.com/hpesc/public/docDisplay?docId=a00115093en_us&page=Shasta_Hardware_Architecture.html}
Accessed 26 March 2025
\end{botherref}
\endbibitem

\bibitem[\protect\citeauthoryear{Roweth et~al.}{2022}]{cug2022duncan}
\begin{bchapter}
\bauthor{\bsnm{Roweth}, \binits{D.}},
\bauthor{\bsnm{Faanes}, \binits{G.}},
\bauthor{\bsnm{Terpstra}, \binits{M.}},
\bauthor{\bsnm{Treger}, \binits{J.}}:
\bctitle{Hpe slingshot launched into network space}.
In: \bbtitle{2022 Cray Users Group}
(\byear{2022}).
\burl{\url{https://cug.org/proceedings/cug2022_proceedings/includes/files/pap121s2-file1.pdf}}
\end{bchapter}
\endbibitem

\bibitem[\protect\citeauthoryear{}{}]{hpe_ex}
\begin{botherref}
{HPE Cray EX Cabinet}.
\url{https://www.hpe.com/psnow/doc/a50002389enw}
Accessed 26 March 2025
\end{botherref}
\endbibitem

\bibitem[\protect\citeauthoryear{Kim et~al.}{2008}]{4556717}
\begin{bchapter}
\bauthor{\bsnm{Kim}, \binits{J.}},
\bauthor{\bsnm{Dally}, \binits{W.J.}},
\bauthor{\bsnm{Scott}, \binits{S.}},
\bauthor{\bsnm{Abts}, \binits{D.}}:
\bctitle{Technology-driven, highly-scalable dragonfly topology}.
In: \bbtitle{2008 International Symposium on Computer Architecture},
pp. \bfpage{77}--\blpage{88}
(\byear{2008}).
\doiurl{10.1109/ISCA.2008.19}
\end{bchapter}
\endbibitem

\bibitem[\protect\citeauthoryear{Latham et~al.}{2025}]{Latham2025DAOS}
\begin{bchapter}
\bauthor{\bsnm{Latham}, \binits{R.}},
\bauthor{\bsnm{Ross}, \binits{R.B.}},
\bauthor{\bsnm{Carns}, \binits{P.}},
\bauthor{\bsnm{Snyder}, \binits{S.}},
\bauthor{\bsnm{Harms}, \binits{K.}},
\bauthor{\bsnm{Velusamy}, \binits{K.}},
\bauthor{\bsnm{Coffman}, \binits{P.}},
\bauthor{\bsnm{McPheeters}, \binits{G.}}:
\bctitle{Initial experiences with {DAOS} object storage on {Aurora}}.
In: \bbtitle{Proceedings of the SC '24 Workshops of the International Conference on High Performance Computing, Network, Storage, and Analysis}.
\bsertitle{SC-W '24},
pp. \bfpage{1304}--\blpage{1310}.
\bpublisher{IEEE Computer Society},
\blocation{Los Alamitos, CA, USA}
(\byear{2025}).
\doiurl{10.1109/SCW63240.2024.00171}
\end{bchapter}
\endbibitem

\bibitem[\protect\citeauthoryear{}{}]{IO500}
\begin{botherref}
{IO500}.
\url{https://io500.org}
Accessed 26 March 2025
\end{botherref}
\endbibitem

\bibitem[\protect\citeauthoryear{Kulyavtsev et~al.}{}]{eagle}
\begin{botherref}
\oauthor{\bsnm{Kulyavtsev}, \binits{A.}},
\oauthor{\bsnm{Cherry}, \binits{A.}},
\oauthor{\bsnm{Harms}, \binits{K.}},
\oauthor{\bsnm{McPheeters}, \binits{G.}}:
{Community Storage using Lustre and Globus Sharing}.
\url{https://www.opensfs.org/wp-content/uploads/LUG23-ALCF-Community-FS.v1.pdf}
Accessed 26 March 2025
\end{botherref}
\endbibitem

\bibitem[\protect\citeauthoryear{}{}]{oneapi}
\begin{botherref}
{Intel oneAPI}.
\url{https://www.intel.com/content/www/us/en/developer/tools/oneapi/overview.html}
Accessed 26 March 2025
\end{botherref}
\endbibitem

\bibitem[\protect\citeauthoryear{}{}]{MKL}
\begin{botherref}
{Intel(R) oneAPI Math Kernel Library (oneMKL)}.
\url{https://www.intel.com/content/www/us/en/developer/tools/oneapi/onemkl.html}
Accessed 26 March 2025
\end{botherref}
\endbibitem

\bibitem[\protect\citeauthoryear{Guo et~al.}{2025}]{doi:10.1177/10943420241311608}
\begin{barticle}
\bauthor{\bsnm{Guo}, \binits{Y.}},
\bauthor{\bsnm{Raffenetti}, \binits{K.}},
\bauthor{\bsnm{Zhou}, \binits{H.}},
\bauthor{\bsnm{Balaji}, \binits{P.}},
\bauthor{\bsnm{Si}, \binits{M.}},
\bauthor{\bsnm{Amer}, \binits{A.}},
\bauthor{\bsnm{Iwasaki}, \binits{S.}},
\bauthor{\bsnm{Seo}, \binits{S.}},
\bauthor{\bsnm{Congiu}, \binits{G.}},
\bauthor{\bsnm{Latham}, \binits{R.}},
\bauthor{\bsnm{Oden}, \binits{L.}},
\bauthor{\bsnm{Gillis}, \binits{T.}},
\bauthor{\bsnm{Zambre}, \binits{R.}},
\bauthor{\bsnm{Ouyang}, \binits{K.}},
\bauthor{\bsnm{Archer}, \binits{C.}},
\bauthor{\bsnm{Bland}, \binits{W.}},
\bauthor{\bsnm{Jose}, \binits{J.}},
\bauthor{\bsnm{Sur}, \binits{S.}},
\bauthor{\bsnm{Fujita}, \binits{H.}},
\bauthor{\bsnm{Durnov}, \binits{D.}},
\bauthor{\bsnm{Chuvelev}, \binits{M.}},
\bauthor{\bsnm{Zheng}, \binits{G.}},
\bauthor{\bsnm{Brooks}, \binits{A.}},
\bauthor{\bsnm{Thapaliya}, \binits{S.}},
\bauthor{\bsnm{Doodi}, \binits{T.}},
\bauthor{\bsnm{Garzaran}, \binits{M.}},
\bauthor{\bsnm{Oyanagi}, \binits{S.}},
\bauthor{\bsnm{Snir}, \binits{M.}},
\bauthor{\bsnm{Thakur}, \binits{R.}}:
\batitle{{Preparing MPICH for exascale}}.
\bjtitle{The International Journal of High Performance Computing Applications}
\bvolume{39}(\bissue{2}),
\bfpage{283}--\blpage{305}
(\byear{2025})
\doiurl{10.1177/10943420241311608}
\end{barticle}
\endbibitem

\bibitem[\protect\citeauthoryear{Raffenetti et~al.}{2017}]{mpich-sc17}
\begin{bchapter}
\bauthor{\bsnm{Raffenetti}, \binits{K.}},
\bauthor{\bsnm{Amer}, \binits{A.}},
\bauthor{\bsnm{Oden}, \binits{L.}},
\bauthor{\bsnm{Archer}, \binits{C.}},
\bauthor{\bsnm{Bland}, \binits{W.}},
\bauthor{\bsnm{Fujita}, \binits{H.}},
\bauthor{\bsnm{Guo}, \binits{Y.}},
\bauthor{\bsnm{Janjusic}, \binits{T.}},
\bauthor{\bsnm{Durnov}, \binits{D.}},
\bauthor{\bsnm{Blocksome}, \binits{M.}},
\bauthor{\bsnm{Si}, \binits{M.}},
\bauthor{\bsnm{Seo}, \binits{S.}},
\bauthor{\bsnm{Langer}, \binits{A.}},
\bauthor{\bsnm{Zheng}, \binits{G.}},
\bauthor{\bsnm{Takagi}, \binits{M.}},
\bauthor{\bsnm{Coffman}, \binits{P.}},
\bauthor{\bsnm{Jose}, \binits{J.}},
\bauthor{\bsnm{Sur}, \binits{S.}},
\bauthor{\bsnm{Sannikov}, \binits{A.}},
\bauthor{\bsnm{Oblomov}, \binits{S.}},
\bauthor{\bsnm{Chuvelev}, \binits{M.}},
\bauthor{\bsnm{Hatanaka}, \binits{M.}},
\bauthor{\bsnm{Zhao}, \binits{X.}},
\bauthor{\bsnm{Fischer}, \binits{P.}},
\bauthor{\bsnm{Rathnayake}, \binits{T.}},
\bauthor{\bsnm{Otten}, \binits{M.}},
\bauthor{\bsnm{Min}, \binits{M.}},
\bauthor{\bsnm{Balaji}, \binits{P.}}:
\bctitle{Why is mpi so slow? analyzing the fundamental limits in implementing mpi-3.1}.
In: \bbtitle{Proceedings of the International Conference for High Performance Computing, Networking, Storage and Analysis}.
\bsertitle{SC '17}.
\bpublisher{Association for Computing Machinery},
\blocation{New York, NY, USA}
(\byear{2017}).
\doiurl{10.1145/3126908.3126963}
\end{bchapter}
\endbibitem

\bibitem[\protect\citeauthoryear{Doodi et~al.}{2021}]{mpich-high-radix21}
\begin{bchapter}
\bauthor{\bsnm{Doodi}, \binits{T.}},
\bauthor{\bsnm{Islam}, \binits{N.}},
\bauthor{\bsnm{Zheng}, \binits{G.}},
\bauthor{\bsnm{Kalidas}, \binits{R.}},
\bauthor{\bsnm{Langer}, \binits{A.}},
\bauthor{\bsnm{Maria}, \binits{G.}}:
\bctitle{High radix collective algorithms}.
In: \bbtitle{Proceedings of EuroMPI}
(\byear{2021})
\end{bchapter}
\endbibitem

\bibitem[\protect\citeauthoryear{Zambre et~al.}{2020}]{mpich-threading}
\begin{bchapter}
\bauthor{\bsnm{Zambre}, \binits{R.}},
\bauthor{\bsnm{Chandramowliswharan}, \binits{A.}},
\bauthor{\bsnm{Balaji}, \binits{P.}}:
\bctitle{How i learned to stop worrying about user-visible endpoints and love mpi}.
In: \bbtitle{Proceedings of the 34th ACM International Conference on Supercomputing}.
\bsertitle{ICS '20}.
\bpublisher{Association for Computing Machinery},
\blocation{New York, NY, USA}
(\byear{2020}).
\doiurl{10.1145/3392717.3392773}
\end{bchapter}
\endbibitem

\bibitem[\protect\citeauthoryear{}{}]{yaksa}
\begin{botherref}
High-Performance Data Type Engine.
\url{https://www.yaksa.org}
Accessed 9 September 2025
\end{botherref}
\endbibitem

\bibitem[\protect\citeauthoryear{}{}]{pmix}
\begin{botherref}
Application programming interface for exascale systems.
\url{https://pmix.github.io/}
Accessed 26 March 2025
\end{botherref}
\endbibitem

\bibitem[\protect\citeauthoryear{}{}]{pals}
\begin{botherref}
HPE Parallel Application Launch Service (PALS).
\url{https://support.hpe.com/hpesc/public/docDisplay?docId=a00117940en_us&page=Parallel_Application_Launch_Service_PALS.html&docLocale=en_US}
Accessed 26 March 2025
\end{botherref}
\endbibitem

\bibitem[\protect\citeauthoryear{}{}]{openmp}
\begin{botherref}
{OpenMP}.
\url{https://www.openmp.org/}
Accessed 26 March 2025
\end{botherref}
\endbibitem

\bibitem[\protect\citeauthoryear{}{}]{hip}
\begin{botherref}
{HIP documentation}.
\url{https://rocm.docs.amd.com/projects/HIP/en/latest/}
Accessed 26 March 2025
\end{botherref}
\endbibitem

\bibitem[\protect\citeauthoryear{Zhao et~al.}{2023}]{chipStar}
\begin{bchapter}
\bauthor{\bsnm{Zhao}, \binits{J.}},
\bauthor{\bsnm{Bertoni}, \binits{C.}},
\bauthor{\bsnm{Young}, \binits{J.}},
\bauthor{\bsnm{Harms}, \binits{K.}},
\bauthor{\bsnm{Sarkar}, \binits{V.}},
\bauthor{\bsnm{Videau}, \binits{B.}}:
\bctitle{{HIPLZ: Enabling Performance Portability for Exascale Systems}}.
In: \beditor{\bsnm{Singer}, \binits{J.}},
\beditor{\bsnm{Elkhatib}, \binits{Y.}},
\beditor{\bsnm{Blanco~Heras}, \binits{D.}},
\beditor{\bsnm{Diehl}, \binits{P.}},
\beditor{\bsnm{Brown}, \binits{N.}},
\beditor{\bsnm{Ilic}, \binits{A.}} (eds.)
\bbtitle{Euro-Par 2022: Parallel Processing Workshops},
pp. \bfpage{197}--\blpage{210}.
\bpublisher{Springer},
\blocation{Cham}
(\byear{2023})
\end{bchapter}
\endbibitem

\bibitem[\protect\citeauthoryear{Babej and J\"{a}\"{a}skel\"{a}inen}{2020}]{chipStar_cl}
\begin{bchapter}
\bauthor{\bsnm{Babej}, \binits{M.}},
\bauthor{\bsnm{J\"{a}\"{a}skel\"{a}inen}, \binits{P.}}:
\bctitle{{HIPCL: Tool for Porting CUDA Applications to Advanced OpenCL Platforms Through HIP}}.
In: \bbtitle{Proceedings of the International Workshop on OpenCL}.
\bsertitle{IWOCL '20}.
\bpublisher{Association for Computing Machinery},
\blocation{New York, NY, USA}
(\byear{2020}).
\doiurl{10.1145/3388333.3388641}
\end{bchapter}
\endbibitem

\bibitem[\protect\citeauthoryear{Arndt et~al.}{2024}]{kokkos_sycl}
\begin{bchapter}
\bauthor{\bsnm{Arndt}, \binits{D.}},
\bauthor{\bsnm{Lebrun-Grandie}, \binits{D.}},
\bauthor{\bsnm{Trott}, \binits{C.}}:
\bctitle{Experiences with implementing kokkos’ sycl backend}.
In: \bbtitle{Proceedings of the 12th International Workshop on OpenCL and SYCL}.
\bsertitle{IWOCL '24}.
\bpublisher{Association for Computing Machinery},
\blocation{New York, NY, USA}
(\byear{2024}).
\doiurl{10.1145/3648115.3648118}
\end{bchapter}
\endbibitem

\bibitem[\protect\citeauthoryear{Homerding et~al.}{2024}]{raja_sycl}
\begin{bchapter}
\bauthor{\bsnm{Homerding}, \binits{B.}},
\bauthor{\bsnm{Vargas}, \binits{A.}},
\bauthor{\bsnm{Scogland}, \binits{T.}},
\bauthor{\bsnm{Chen}, \binits{R.}},
\bauthor{\bsnm{Davis}, \binits{M.}},
\bauthor{\bsnm{Hornung}, \binits{R.}}:
\bctitle{{Enabling RAJA on Intel GPUs with SYCL}}.
In: \bbtitle{Proceedings of the 12th International Workshop on OpenCL and SYCL}.
\bsertitle{IWOCL '24}.
\bpublisher{Association for Computing Machinery},
\blocation{New York, NY, USA}
(\byear{2024}).
\doiurl{10.1145/3648115.3648131}
\end{bchapter}
\endbibitem

\bibitem[\protect\citeauthoryear{}{}]{unittrace}
\begin{botherref}
Intel unittraceß code repository.
\url{https://github.com/intel/pti-gpu/tree/master/tools/unitrace}
Accessed 26 March 2025
\end{botherref}
\endbibitem

\bibitem[\protect\citeauthoryear{Bekele et~al.}{2025}]{bekele2025thapi}
\begin{botherref}
\oauthor{\bsnm{Bekele}, \binits{S.}},
\oauthor{\bsnm{Vivas}, \binits{A.}},
\oauthor{\bsnm{Applencourt}, \binits{T.}},
\oauthor{\bsnm{Muralidharan}, \binits{S.}},
\oauthor{\bsnm{Allen}, \binits{B.}},
\oauthor{\bsnm{Yoshiiinst}, \binits{K.}},
\oauthor{\bsnm{Perarnau}, \binits{S.}},
\oauthor{\bsnm{Videau}, \binits{B.}}:
{THAPI: Tracing Heterogeneous APIs}
(2025)
\end{botherref}
\endbibitem

\bibitem[\protect\citeauthoryear{Eastep et~al.}{2017}]{geopm}
\begin{bchapter}
\bauthor{\bsnm{Eastep}, \binits{J.}},
\bauthor{\bsnm{Sylvester}, \binits{S.}},
\bauthor{\bsnm{Cantalupo}, \binits{C.}},
\bauthor{\bsnm{Geltz}, \binits{B.}},
\bauthor{\bsnm{Ardanaz}, \binits{F.}},
\bauthor{\bsnm{Al-Rawi}, \binits{A.}},
\bauthor{\bsnm{Livingston}, \binits{K.}},
\bauthor{\bsnm{Keceli}, \binits{F.}},
\bauthor{\bsnm{Maiterth}, \binits{M.}},
\bauthor{\bsnm{Jana}, \binits{S.}}:
\bctitle{{Global Extensible Open Power Manager: A Vehicle for HPC Community Collaboration on Co-Designed Energy Management Solutions}}.
In: \bbtitle{High Performance Computing: 32nd International Conference, ISC High Performance 2017, Frankfurt, Germany, June 18–22, 2017, Proceedings},
pp. \bfpage{394}--\blpage{412}.
\bpublisher{Springer},
\blocation{Berlin, Heidelberg}
(\byear{2017}).
\doiurl{10.1007/978-3-319-58667-0_21}
\end{bchapter}
\endbibitem

\bibitem[\protect\citeauthoryear{}{}]{oneCCL}
\begin{botherref}
oneCCL code repository.
\url{https://github.com/uxlfoundation/oneCCL}
Accessed 26 March 2025
\end{botherref}
\endbibitem

\bibitem[\protect\citeauthoryear{}{}]{OpenVINO}
\begin{botherref}
OpenVINO project code repository.
\url{https://github.com/openvinotoolkit/openvino}
Accessed 26 March 2025
\end{botherref}
\endbibitem

\bibitem[\protect\citeauthoryear{}{}]{vllm}
\begin{botherref}
vLLM project code repository.
\url{https://github.com/vllm-project/vllm}
Accessed 26 March 2025
\end{botherref}
\endbibitem

\bibitem[\protect\citeauthoryear{}{}]{SGLang}
\begin{botherref}
SGLang project code repository.
\url{https://github.com/sgl-project/sglang}
Accessed 26 March 2025
\end{botherref}
\endbibitem

\bibitem[\protect\citeauthoryear{}{}]{HFT}
\begin{botherref}
HuggingFace Transformers library.
\url{https://huggingface.co/docs/transformers/en/index}
Accessed 26 March 2025
\end{botherref}
\endbibitem

\bibitem[\protect\citeauthoryear{}{}]{TRL}
\begin{botherref}
Transformers re-inforcement library code repository.
\url{https://github.com/huggingface/trl}
Accessed 26 March 2025
\end{botherref}
\endbibitem

\bibitem[\protect\citeauthoryear{}{}]{Accelerate}
\begin{botherref}
Accelerate Training library code repository.
\url{https://github.com/huggingface/accelerate}
Accessed 26 March 2025
\end{botherref}
\endbibitem

\bibitem[\protect\citeauthoryear{}{}]{TorchTune}
\begin{botherref}
PyTorch Post-training Library code repository.
\url{https://github.com/pytorch/torchtune}
Accessed 26 March 2025
\end{botherref}
\endbibitem

\bibitem[\protect\citeauthoryear{}{}]{openXLA}
\begin{botherref}
Intel Extensions for OpenXLA code repository.
\url{https://github.com/intel/intel-extension-for-openxla}
Accessed 26 March 2025
\end{botherref}
\endbibitem

\bibitem[\protect\citeauthoryear{}{}]{DPEP}
\begin{botherref}
Data Parallel Extensions for Python code repository.
\url{https://github.com/IntelPython/DPEP}
Accessed 26 March 2025
\end{botherref}
\endbibitem

\bibitem[\protect\citeauthoryear{}{}]{dpnp}
\begin{botherref}
Data Parallel Extension for NumPy code repository.
\url{https://github.com/IntelPython/dpnp}
Accessed 26 March 2025
\end{botherref}
\endbibitem

\bibitem[\protect\citeauthoryear{}{}]{dpctl}
\begin{botherref}
Data Parallel Control code repository.
\url{https://github.com/IntelPython/dpctl}
Accessed 26 March 2025
\end{botherref}
\endbibitem

\bibitem[\protect\citeauthoryear{}{}]{dpex}
\begin{botherref}
Data Parallel Control code repository.
\url{https://github.com/IntelPython/numba-dpex}
Accessed 26 March 2025
\end{botherref}
\endbibitem

\bibitem[\protect\citeauthoryear{}{}]{DLPack}
\begin{botherref}
DLPack Tutorial.
\url{https://intel.github.io/intel-extension-for-pytorch/xpu/latest/tutorials/features/DLPack.html}
Accessed 26 March 2025
\end{botherref}
\endbibitem

\bibitem[\protect\citeauthoryear{}{}]{onedal}
\begin{botherref}
oneAPI Data Analytics Library code repository.
\url{https://github.com/uxlfoundation/oneDAL}
Accessed 26 March 2025
\end{botherref}
\endbibitem

\bibitem[\protect\citeauthoryear{}{}]{Copper}
\begin{botherref}
Copper Library code repository.
\url{https://github.com/argonne-lcf/copper}
Accessed 26 March 2025
\end{botherref}
\endbibitem

\bibitem[\protect\citeauthoryear{Lewis et~al.}{2024}]{Copper_SC24}
\begin{bchapter}
\bauthor{\bsnm{Lewis}, \binits{N.}},
\bauthor{\bsnm{Velusamy}, \binits{K.}},
\bauthor{\bsnm{Harms}, \binits{K.}},
\bauthor{\bsnm{Zheng}, \binits{H.}}:
\bctitle{Copper: Cooperative caching layer for scalable data loading in exascale supercomputers}.
In: \bbtitle{SC24-W: Workshops of the International Conference for High Performance Computing, Networking, Storage and Analysis},
pp. \bfpage{1320}--\blpage{1329}
(\byear{2024}).
\doiurl{10.1109/SCW63240.2024.00173}
\end{bchapter}
\endbibitem

\bibitem[\protect\citeauthoryear{Wald et~al.}{2017}]{7539599}
\begin{barticle}
\bauthor{\bsnm{Wald}, \binits{I.}},
\bauthor{\bsnm{Johnson}, \binits{G.}},
\bauthor{\bsnm{Amstutz}, \binits{J.}},
\bauthor{\bsnm{Brownlee}, \binits{C.}},
\bauthor{\bsnm{Knoll}, \binits{A.}},
\bauthor{\bsnm{Jeffers}, \binits{J.}},
\bauthor{\bsnm{Günther}, \binits{J.}},
\bauthor{\bsnm{Navratil}, \binits{P.}}:
\batitle{Ospray - a cpu ray tracing framework for scientific visualization}.
\bjtitle{IEEE Transactions on Visualization and Computer Graphics}
\bvolume{23}(\bissue{1}),
\bfpage{931}--\blpage{940}
(\byear{2017})
\doiurl{10.1109/TVCG.2016.2599041}
\end{barticle}
\endbibitem

\bibitem[\protect\citeauthoryear{Moreland et~al.}{2024}]{moreland2024visualization}
\begin{barticle}
\bauthor{\bsnm{Moreland}, \binits{K.}},
\bauthor{\bsnm{Athawale}, \binits{T.M.}},
\bauthor{\bsnm{Bolea}, \binits{V.}},
\bauthor{\bsnm{Bolstad}, \binits{M.}},
\bauthor{\bsnm{Brugger}, \binits{E.}},
\bauthor{\bsnm{Childs}, \binits{H.}},
\bauthor{\bsnm{Huebl}, \binits{A.}},
\bauthor{\bsnm{Lo}, \binits{L.-T.}},
\bauthor{\bsnm{Geveci}, \binits{B.}},
\bauthor{\bsnm{Marsaglia}, \binits{N.}}, \betal:
\batitle{Visualization at exascale: Making it all work with vtk-m}.
\bjtitle{The International Journal of High Performance Computing Applications}
\bvolume{38}(\bissue{5}),
\bfpage{508}--\blpage{526}
(\byear{2024})
\end{barticle}
\endbibitem

\bibitem[\protect\citeauthoryear{}{}]{ESP}
\begin{botherref}
ALCF Early Science Program web page.
ALCF Early Science Program web page.
\url{https://www.alcf.anl.gov/science/early-science-program}
Accessed 26 March 2025
\end{botherref}
\endbibitem

\bibitem[\protect\citeauthoryear{}{}]{ECP}
\begin{botherref}
Exascale Computing Project Webpage.
\url{https://www.exascaleproject.org/}
Accessed 26 March 2025
\end{botherref}
\endbibitem

\bibitem[\protect\citeauthoryear{Bagusetty et~al.}{2022}]{nwchemex_sycl}
\begin{bchapter}
\bauthor{\bsnm{Bagusetty}, \binits{A.}},
\bauthor{\bsnm{Panyala}, \binits{A.}},
\bauthor{\bsnm{Brown}, \binits{G.}},
\bauthor{\bsnm{Kirk}, \binits{J.}}:
\bctitle{{Towards Cross-Platform Portability of Coupled-Cluster Methods with Perturbative Triples using SYCL}}.
In: \bbtitle{2022 IEEE/ACM International Workshop on Performance, Portability and Productivity in HPC (P3HPC)},
pp. \bfpage{81}--\blpage{88}
(\byear{2022}).
\doiurl{10.1109/P3HPC56579.2022.00013}
\end{bchapter}
\endbibitem

\bibitem[\protect\citeauthoryear{Mutlu et~al.}{2023}]{nwchemex_tamm}
\begin{botherref}
\oauthor{\bsnm{Mutlu}, \binits{E.}},
\oauthor{\bsnm{Panyala}, \binits{A.}},
\oauthor{\bsnm{Gawande}, \binits{N.}},
\oauthor{\bsnm{Bagusetty}, \binits{A.}},
\oauthor{\bsnm{Glabe}, \binits{J.}},
\oauthor{\bsnm{Kim}, \binits{J.}},
\oauthor{\bsnm{Kowalski}, \binits{K.}},
\oauthor{\bsnm{Bauman}, \binits{N.P.}},
\oauthor{\bsnm{Peng}, \binits{B.}},
\oauthor{\bsnm{Pathak}, \binits{H.}}, et al.:
{TAMM: Tensor algebra for many-body methods}.
The Journal of Chemical Physics
\textbf{159}(2)
(2023)
\end{botherref}
\endbibitem

\bibitem[\protect\citeauthoryear{Bertoni et~al.}{2025}]{gemm_benchmarking}
\begin{bchapter}
\bauthor{\bsnm{Bertoni}, \binits{C.}},
\bauthor{\bsnm{Applencourt}, \binits{T.}},
\bauthor{\bsnm{Gao}, \binits{L.}},
\bauthor{\bsnm{Leggett}, \binits{T.}}:
\bctitle{Millions of matrix-multiplications: Gemm variations on aurora}.
In: \bbtitle{2025 IEEE International Parallel and Distributed Processing Symposium Workshops (IPDPSW)}
(\byear{2025})
\end{bchapter}
\endbibitem

\bibitem[\protect\citeauthoryear{}{}]{HPLMxP}
\begin{botherref}
{HPL-MxP}.
\url{https://hpl-mxp.org/results.md}
Accessed 26 March 2025
\end{botherref}
\endbibitem

\bibitem[\protect\citeauthoryear{}{}]{HPCG}
\begin{botherref}
{HPCG}.
\url{http://www.hpcg-benchmark.org/}
Accessed 26 March 2025
\end{botherref}
\endbibitem

\bibitem[\protect\citeauthoryear{}{}]{GRAPH500}
\begin{botherref}
Graph500.
\url{https://graph500.org}
Accessed 26 March 2025
\end{botherref}
\endbibitem

\bibitem[\protect\citeauthoryear{Mills et~al.}{2021}]{petsc-paper}
\begin{bchapter}
\bauthor{\bsnm{Mills}, \binits{R.T.}},
\bauthor{\bsnm{Adams}, \binits{M.F.}},
\bauthor{\bsnm{Balay}, \binits{S.}},
\bauthor{\bsnm{Brown}, \binits{J.}},
\bauthor{\bsnm{Dener}, \binits{A.}},
\bauthor{\bsnm{Knepley}, \binits{M.}},
\bauthor{\bsnm{Kruger}, \binits{S.E.}},
\bauthor{\bsnm{Morgan}, \binits{H.}},
\bauthor{\bsnm{Munson}, \binits{T.}},
\bauthor{\bsnm{Rupp}, \binits{K.}},
\bauthor{\bsnm{Smith}, \binits{B.F.}},
\bauthor{\bsnm{Zampini}, \binits{S.}},
\bauthor{\bsnm{Zhang}, \binits{H.}},
\bauthor{\bsnm{Zhang}, \binits{J.}}:
\bctitle{Toward performance-portable petsc for gpu-based exascale systems}.
In: \bbtitle{Parallel Computing},
vol. \bseriesno{108}
(\byear{2021}).
\doiurl{10.1016/j.parco.2021.102831}
\end{bchapter}
\endbibitem

\bibitem[\protect\citeauthoryear{Thakur et~al.}{2005}]{thakur_collectives}
\begin{barticle}
\bauthor{\bsnm{Thakur}, \binits{R.}},
\bauthor{\bsnm{Rabenseifner}, \binits{R.}},
\bauthor{\bsnm{Gropp}, \binits{W.}}:
\batitle{Optimization of collective communication operations in mpich}.
\bjtitle{The International Journal of High Performance Computing Applications}
\bvolume{19}(\bissue{1}),
\bfpage{49}--\blpage{66}
(\byear{2005})
\doiurl{10.1177/1094342005051521}
\end{barticle}
\endbibitem

\bibitem[\protect\citeauthoryear{Cui et~al.}{2025}]{cui25arxiv}
\begin{botherref}
\oauthor{\bsnm{Cui}, \binits{S.}},
\oauthor{\bsnm{Patke}, \binits{A.}},
\oauthor{\bsnm{Chen}, \binits{Z.}},
\oauthor{\bsnm{Ranjan}, \binits{A.}},
\oauthor{\bsnm{Nguyen}, \binits{H.}},
\oauthor{\bsnm{Cao}, \binits{P.}},
\oauthor{\bsnm{Jha}, \binits{S.}},
\oauthor{\bsnm{Bode}, \binits{B.}},
\oauthor{\bsnm{Bauer}, \binits{G.}},
\oauthor{\bsnm{Narayanaswami}, \binits{C.}},
\oauthor{\bsnm{Sow}, \binits{D.}},
\oauthor{\bsnm{Martino}, \binits{C.D.}},
\oauthor{\bsnm{Kalbarczyk}, \binits{Z.T.}},
\oauthor{\bsnm{Iyer}, \binits{R.K.}}:
{Characterizing GPU Resilience and Impact on AI/HPC Systems}
(2025).
\url{https://arxiv.org/abs/2503.11901}
\end{botherref}
\endbibitem

\bibitem[\protect\citeauthoryear{Team}{2024}]{meta24arxiv}
\begin{botherref}
\oauthor{\bsnm{Team}, \binits{M.L.}}:
{The Llama 3 Herd of Models}
(2024).
\url{https://arxiv.org/abs/2407.21783}
\end{botherref}
\endbibitem

\bibitem[\protect\citeauthoryear{Kokolis et~al.}{2025}]{kokolis25arxiv}
\begin{botherref}
\oauthor{\bsnm{Kokolis}, \binits{A.}},
\oauthor{\bsnm{Kuchnik}, \binits{M.}},
\oauthor{\bsnm{Hoffman}, \binits{J.}},
\oauthor{\bsnm{Kumar}, \binits{A.}},
\oauthor{\bsnm{Malani}, \binits{P.}},
\oauthor{\bsnm{Ma}, \binits{F.}},
\oauthor{\bsnm{DeVito}, \binits{Z.}},
\oauthor{\bsnm{Sengupta}, \binits{S.}},
\oauthor{\bsnm{Saladi}, \binits{K.}},
\oauthor{\bsnm{Wu}, \binits{C.-J.}}:
Revisiting Reliability in Large-Scale Machine Learning Research Clusters
(2025).
\url{https://arxiv.org/abs/2410.21680}
\end{botherref}
\endbibitem

\bibitem[\protect\citeauthoryear{Gerofi}{2025}]{bgerofi25MW}
\begin{botherref}
\oauthor{\bsnm{Gerofi}, \binits{B.}}:
{Fine-grained Automated Failure Management on Extreme-Scale GPU Accelerated Systems}
(2025).
\url{https://multicore.world/speakers-2025/balazs-gerofi/}
Accessed 26 March 2025
\end{botherref}
\endbibitem

\bibitem[\protect\citeauthoryear{Kates-Harbeck et~al.}{2019}]{FusionDL}
\begin{barticle}
\bauthor{\bsnm{Kates-Harbeck}, \binits{J.}},
\bauthor{\bsnm{Svyatkovskiy}, \binits{A.}},
\bauthor{\bsnm{Tang}, \binits{W.}}:
\batitle{Predicting disruptive instabilities in controlled fusion plasmas through deep learning}.
\bjtitle{Nature}
\bvolume{568}(\bissue{7753}),
\bfpage{526}--\blpage{531}
(\byear{2019})
\doiurl{10.1038/s41586-019-1116-4}
\end{barticle}
\endbibitem

\end{thebibliography}

\end{document}